\documentstyle[12pt,epsf]{article}

\begin{titlepage}                                                                          
\title{ 
{\bf          Two-loop  renormalization  group }\\      
{\bf        restrictions on  the  standard  model }\\
{\bf        and the fourth chiral family }\\[3 ex]
}
                                                                     
\author{ Yu.~F.~Pirogov \and O.~V.~Zenin\\[2.5 ex]                                        
{\it Institute for High Energy Physics,}\\
{\it Protvino, Moscow Region, Russia}\\[0.5ex]
{\it  Moscow  Institute  of  Physics  and  Technology,}\\
{\it  Dolgoprudny,  Moscow  Region, Russia }}
\date{}

\begin{document}
\maketitle
\thispagestyle{empty}

\newpage
\abstract{
\noindent
In the framework of the two-loop renormalization group, the global
profile of the Standard Model (SM) in its full parameter space is
investigated up to the scale of
the gauge singularity. The critical Higgs masses bordering the strong 
coupling, unstable and safe regions are explicitly found.
Restrictions on the Higgs boson mass as a function of a cutoff scale
are obtained from the stability of electroweak vacuum and from the
absence of the strong coupling both in the Higgs and Yukawa sectors. 
The cutoff being equal to the Planck scale requires
the Higgs mass to be $M_H = (161.3 \pm 20.6)^{+4}_{-10}$ GeV and
$M_H\ge 140.7^{+10}_{-10}$ GeV, where the $M_H$ corridor is the
theoretical one and the errors are due to the top mass
uncertainty. 

The SM two-loop $\beta$ functions are generalized to the massive
neutrino case. Modification of the two-loop global profile of the SM
extended by one new chiral
family is studied, and bounds on the masses of the family are
derived. The requirement of self-consistency of the perturbative SM
as an underlying theory up to the Planck or GUT scale excludes
the fourth chiral family. But as an effective theory, the SM allows
the heavy chiral family with the mass up to 250 GeV depending on the
Higgs mass and the cutoff scale.  Under precision experiment
restriction $M_H\leq 200$~GeV, the fourth chiral family,  taken
alone, is excluded. Nevertheless  a pair of the chiral families
constituting  the vector-like one  could still exist.
}
\thispagestyle{empty}
\end{titlepage}

\addtocounter{page}{-1}

\section{ Introduction }

The renormalization group (RG) study of a field theory (for the
review see, e.g., Refs.~\cite{peterman,shirkov}) enables one
to understand in
grosso the structure of the theory as a function of a characteristic
energy scale. Of special interest are the cases when self-consistency
of the theory is under danger of violation. They may signal either
the breakdown of the perturbative validity or/and the onset of a
``new physics''.

There are two problems of the kind in the Standard Model (SM).
First, it encounters when some of the running couplings tend to blow
up at the finite scales. The well-known examples are, e.g., the
Landau singularity
in QED (more generally, in any Abelian gauge theory $U(1)$) or in the
$\phi^4$ scalar theory. In the latter case, the problem was known for
a long time as the triviality problem (for the review of triviality
arguments see,  e.g.,
Ref.~\cite{lindner1} and references therein). 
Technically, it can be hoped to be solved by an
improvement of the perturbative series or by development of the
strong coupling methods.  But more probably it has a physical
origin, and it could be solved  eventually by the more complete
theory  which should result effectively in a physical cutoff (for an
example see,  e.g., Ref.~\cite{petronzio}). In particular, this
problem was invoked  to justify the technicolor as a substitute
for the heavy Higgs boson.  
 
Second, the problem occurs when a running coupling leaves  the
physical
region at some finite scale. In the SM, this happens when the Higgs
quartic effective coupling becomes negative, indicating the absence
of a ground state in the quantum theory. It is the so-called
electroweak vacuum stability problem (for a review  see,
e.g., Ref.~\cite{stability}). It is a real problem of the quantum
field theory because this phenomenon takes place in the realm of the
perturbative validity. In the framework of the SM, the light Higgs
bosons resulting in the unstable electroweak vacuum should be
forbidden. On the other hand, if this does happen
some new scalar bosons beyond the SM will be required to stabilize
the
vacuum. Otherwise, the light composite Higgs 
with the compositeness scale corresponding to the scale of the
stability breakdown might be a natural solution.

The SM self-consistency study in the framework of the one-loop RG
and restrictions thereof on the SM heavy particles, the Higgs boson
and the top quark, was undertaken in
Refs.~\cite{lindner1,cabibbo,beg}. A generalization to
the two-loop level was given in Refs.~\cite{lindner2}--\cite{lee}.
The one-loop RG restrictions on a new heavy chiral family were
studied in Ref.~\cite{novikov}, and that on the vector-like family
were investigated in Ref.~\cite{zheng}.

  The aim of our present study is twofold. First, we investigate
the two-loop RG global profile of the SM in its parameter space at
all conceivable scales. In particular, we refine the RG 
restrictions on the Higgs mass in the light of the now accurately
known top mass and its uncertainty. This provides us with the
background required for the RG study of possible extensions to the
SM. Second, we generalize  to the two-loop level  
the RG study of the SM extended by the fourth chiral family,
as well as refine the self-consistency restrictions thereof on the
Higgs and  fourth
family masses. This required in turn a generalization of the SM
two-loop $\beta$ functions to the massive neutrino case, which we
present. In a wider perspective, we consider the problem of what
principally new the fourth heavy chiral family brings in the RG
global profile of the SM.

\section{ Standard Model }

The two-loop $\beta$ functions for a general gauge theory  in the
$\overline{\rm MS}$ renormalization scheme are well-known in the
literature~\cite{machacek} as well as their particular realization
for the SM~\cite{machacek}--\cite{ford2}  (compact summaries for the
SM can also be found in Refs.~\cite{barger,schrempp}). 
They are re-collected in a different form 
in the Appendix A1 with the explicit Yukawa couplings being retained
only for the third family.
In what follows we put just the generic structure of the emerging
one- and two-loop RG differential equations. 

Let $g_{i}$, $ i=1,2,3$, $y_ f$, $\lambda$ and $v$
be the SM gauge couplings, the Yukawa couplings for fermions  
$ f$, the  Higgs self-interaction 
coupling  and the vacuum expectation value (VEV), respectively.
 Then one gets
\begin{eqnarray}
\frac{1}{g_{i}^{3}}\frac{dg_{i}}{d\ln\mu}&=& 
\frac{1}{(4\pi)^{2}} b_{g_{i}}^{(1)} (g_{i'}) + 
\frac{1}{(4\pi)^{4}} b_{g_{i}}^{(2)}(g_{i'}, y_{f'}),
\nonumber\\   
\frac{1}{y_{f}}\frac{dy_{f}}{d\ln\mu}&=&
\frac{1}{(4\pi)^{2}} b_{y_{f}}^{(1)}(g_{i'},y_{f'}) + 
\frac{1}{(4\pi)^{4}} b_{y_{f}}^{(2)}(g_{i'}, y_{f'},\lambda),
\nonumber\\
\frac{d\lambda}{d\ln\mu} &=&
\frac{1}{(4\pi)^{2}} b_{\lambda}^{(1)}(g_{i'},y_{f'},\lambda) + 
\frac{1}{(4\pi)^{4}} b_{\lambda}^{(2)}(g_{i'}, y_{f'}, \lambda),
\nonumber\\
\frac{1}{v}\frac{dv}{d\ln\mu} &=&
\frac{1}{(4\pi)^{2}} b_{v}^{(1)}(g_{i'},y_{f'}) + 
\frac{1}{(4\pi)^{4}} b_{v}^{(2)}(g_{i'}, y_{f'}, \lambda),
\end{eqnarray}
where ~$\mu$~ is a renormalization scale, say in GeV, $b^{(1)}$ and
$b^{(2)}$ are one- and two-loop contributions respectively,
$b_{g_i}^{(1)}$ being in fact the  constants. Under $g_{i'}$ and
$y_{f'}$ there
are understood the sets of all $g_{i}$ and $y_{f}$.
We neglected here for simplicity by the mixing of the Yukawa
couplings  and thus by  the CP violating phase. To state it in
other terms, the diagonal real form of
the Yukawa matrix $Y_{ff'} = y_{f}\delta_{ff'}$ is
implied. 

   The following essential features of the RG system (1) are readily 
ascertained. At one-loop order, one has a kind of the three-level
up-down
hierarchy among the SM couplings, so that  the first three
differential equations in (1) disentangle. One can first find
$g_{i}(\mu)$, then insert them
into  $\beta_{y_{f}}^{(1)}$ and find $y_{f}(\mu)$, and finally put 
$g_{i}(\mu)$, $y_{f}(\mu)$ into the third equation and integrate it.
The solution to the equation for $v(\mu)$ is determined completely by
those for the first three equations, both in one and two
loops.\footnote{For the sake of completeness, the evolution of the
gauge fixing parameter $\xi$ should also be taken into account in the
gauge dependent quantities like $v(\mu)$. To avoid this one can use
the 't~Hooft--Landau gauge $\xi=0$ which is not running.} 

  In two loops, the  RG equations partially entangle with each other
due to a down-up feedback to the neighbour level: from $\lambda$ to
$y_{f}$ and from $y_{f}$ to $g_{i}$. But there is no direct influence
of $\lambda$ on $g_{i}$. It emerges only in three loops. Hence to
completely entangle the RG system  one needs the three-loop SM
$\beta$ functions, which are unknown at present.  Thus we have to
restrict ourselves to the two-loop order. On the other hand, the 
two- and higher-loop contributions to $\beta$ functions, even the
sign including, are known to depend in a multi-coupling theory on the
renormalization scheme \cite{shirkov}. 
Hence the physical meaning of the running couplings becomes
ambiguous, and it is impossible to improve the perturbative RG
analysis of the SM in the scheme-independent way beyond one loop.

We integrated the RG Eq.~(1) numerically for $\mu\ge M_Z$ by the
first-order Runge-Cutta method with the initial conditions at the 
scale $M_{Z}$ taken as
\begin{eqnarray}
\alpha_{1}(M_Z)&=&0.0102,\nonumber\\
\alpha_{2}(M_Z)&=&0.0338,\nonumber\\
\alpha_{3}(M_Z)&=&0.123 
\end{eqnarray}
in accordance with  $\alpha(M_Z)=1/127.90$ and 
$\sin^{2}\theta_{W}(M_Z)=0.2315$~\cite{rpp}.
Our normalizations of the gauge couplings are as follows:
$g_{1}=(5/3)^{1/2}g'$, $g_2\equiv g$ and $g_3\equiv g_S$, with $g'$,
$g$ and $g_S$  being the conventional SM couplings. We choose also
the relations $m_{f}=y_{f}v$  and $m_{H}=\lambda^{1/2} v$  as the
definition of normalization for the Higgs and Yukawa couplings, with
$v=(\sqrt{2}\,G_F)^{-1/2}=246.22$ GeV being the Higgs VEV. Because
the evolution of $v(\mu)$ is gauge dependent we use in what follows
only the gauge independent observable $v\equiv v(M_Z)$.

Besides, we use  at $\mu=M_Z$ the one-loop  matching condition
for the physical $M_f$ and  running  $m_f(\mu)\equiv y_f(\mu)\,v$
masses  of the fermions  $f=q$ and $l$  given by
\begin{equation}
m_f(\mu)=M_f \Big(1+\delta_f^{\mbox{\scriptsize QCD}}(\mu)
+\delta_f^{\mbox{\scriptsize QED}}(\mu)+\delta_f^{(\mbox{\scriptsize
t,\,H})}(\mu)\Big).  
\end{equation}
Here one has
\begin{equation}
\delta_q^{\mbox{\scriptsize QCD}}(\mu) = -
\frac{4}{3}\frac{\alpha_3(\mu)}{\pi}
\Bigl(1+\frac{3}{4}\mbox{ln}\frac{\mu^2}{M_q^2}\Bigr), 
\end{equation}
with $\delta_f^{\mbox{\scriptsize QED}}$ obtained from the last
Eq.\ by
substituting $4/3\,\alpha_3$ by $Q_f^2\alpha$, where $Q_f$ is the
electric charge of the fermion $f$.
The top quark and Higgs boson induced radiative corrections
$\delta_f^{(\mbox{\scriptsize t,\,H})}$ can be found in
Ref.~\cite{kniel} as
\begin{eqnarray} 
\delta_\tau^{(\mbox{\scriptsize t,\,H})}(\mu)&=&
\frac{1}{(4\pi)^2}\left(\frac{M_t}{v}\right)^2
\left(3\,\mbox{ln}\frac{\mu^2}{M_t^2}+\frac{3}{2}+\frac{1}{4}
\frac{M_H^2}{M_t^2} \right), \nonumber\\
\delta_b^{(\mbox{\scriptsize t,\,H})}(\mu)&=&
\frac{1}{(4\pi)^2}\left(\frac{M_t}{v}\right)^2
\left(\frac{3}{2}\,\mbox{ln}\frac{\mu^2}{M_t^2}+
\frac{1}{4}+\frac{1}{4}
\frac{M_H^2}{M_t^2} \right),\nonumber\\
\delta_t^{(\mbox{\scriptsize t,\,H})}(\mu)&=&
\frac{1}{(4\pi)^2}\left(\frac{M_t}{v}\right)^2
\left(\frac{9}{2}\,\mbox{ln}\frac{\mu^2}{M_t^2}+\frac{11}{2}-2\pi
\frac{M_H}{M_t}
\right),
\end{eqnarray}
where  the last line is valid at $(M_H/2 M_t)^2\ll 1$.
Similarly, the initial value $m_H(M_Z)$ is related  with the physical
Higgs mass $M_H$ through the running  mass $m_H(\mu)\equiv
y_H(\mu)\,v$ at the scale $M_Z$, where
\begin{equation}\label{delta_H}
m_H(\mu)=M_H\left(1+\delta_H(\mu)\right). 
\end{equation}
In the limit $(M_H/2 M_t)^2 \ll 1$ one has the following asymptotic
one-loop expression~\cite{kniel,sirlin} 
\begin{eqnarray}
(4\pi)^2 \delta_H(\mu) &=& \,\,\,\,
\left(\frac{M_H}{v}\right)^2
\left(3\, \mbox{ln}\frac{\mu^2}{M_H^2} + 6 -\frac{3\sqrt{3}\pi}{4}
\right)
\nonumber\\
&& + \left(\frac{M_t}{v}\right)^2
     \left(3\, \mbox{ln}\frac{\mu^2}{M_t^2} 
     - 2 + \frac{3}{10} \frac{M_H^2}{ M_t^2} \right)
\nonumber\\
&& + \left(\frac{ M_W}{v}\right)^2
     \left[3\,\mbox{ln}\frac{M_W^2}{\mu^2} + 
    \frac{1}{2} \frac{M_H^2}{ M_W^2}\mbox{ln}\frac{M_H^2}{M_W^2}
     - 5 + 12 \frac{M_W^2}{M_H^2}\right.
\nonumber\\
&&  \phantom{\frac{1}{2}\left(\frac{2 M_Z}{v}\right)^2}
-\left.4\Big(3 \frac{M_W^2}{M_H^2}
          + \frac{1}{4}\frac{M_H^2}{ M_W^2} - 1\Big)\, f
          \Big(\frac{M_W^2}{M_H^2}\Big)\right] 
\nonumber\\
&& + \left(\frac{ M_Z}{v}\right)^2
     \left[\frac{3}{2}\,\mbox{ln}\frac{M_Z^2}{\mu^2} + 
    \frac{1}{4} \frac{m_H^2}{ M_Z^2}\mbox{ln}\frac{M_H^2}{M_Z^2}
     - \frac{5}{2} + 6 \frac{M_Z^2}{M_H^2}\right.
\nonumber\\
&& \phantom{\frac{1}{2}\left(\frac{2 M_Z}{v}\right)^2}- \left. 
2\Big(3 \frac{M_Z^2}{M_H^2}
          +\frac{1}{4} \frac{M_H^2}{ M_Z^2} - 1\Big)\, f
          \Big(\frac{M_Z^2}{M_H^2}\Big)
  \right]\, , 
\end{eqnarray}
where 

\begin{equation} 
f (x) = \cases{
(4 x - 1)^{1/2}\,\mbox{arctg} (4 x - 1)^{-1/2} \, ,\quad\,\, x >
\frac{1}{4} 
\nonumber\cr
\frac{1}{2}\,(1 - 4 x)^{1/2}\,\mbox{ln} 
\frac{1 + (1 - 4 x)^{1/2}}{1 - (1 - 4 x)^{1/2}}\, , 
\qquad\quad x < \frac{1}{4}\,\, .
\cr 
}
\end{equation}

Putting all together one  gets finally at $M_H=150$~GeV:
\begin{eqnarray}
m_\tau(M_Z)&=&1.764~\mbox{GeV},\nonumber\\  
m_b(M_Z)   &=&(4.47\pm 0.50)~ \mbox{GeV},\nonumber\\
m_t(M_Z)   &=&(171.8^{+4.6}_{-4.7})~ \mbox{GeV}.
\end{eqnarray}
The last two values correspond in turn to the physical bottom and
top masses $M_b=(4.5\pm 0.5)$ GeV~\cite{rpp} and  $M_{t}=(175 \pm
5)$ GeV~\cite{top},
respectively. Only errors in the top mass are left as the main
source of the subsequent uncertainties. 
                                                                                                                                                                                                                                                               
As a field theory, SM is legitimate to be pulled to its inner
ultimate limits.
This may help to understand better its structure in the physically
reasonable region  $\mu<M_{Pl}$, $M_{Pl}\simeq 10^{19}$ GeV, which is
to be
considered more seriously. So all the subsequent numerical results
are obtained at all allowed $\mu$ with the exact two-loop $\beta$
functions.
Most of the terms in the latter ones proved to be crucial for the
quantitative
evolution of couplings in the physical $\mu$ region up to the Planck
scale. But for the qualitative analysis of the SM  RG solutions at
extremely high
$\mu$, $\mu\gg M_{Pl}$, we retain in the coefficients of the $\beta$
functions given below only the most representative terms.

To estimate the dependence of the results on the loop order and to
pick out regions where perturbation theory may be more reliably
trusted, we present
both the one- and two-loop results.  They are shown in Figs.~1--5.
Let us discuss them in turn for the gauge, Yukawa and Higgs sectors
of the~SM.

\subsection*{({\bf\em i}) Gauge  sector}

Fig.~1  shows the running with $\mu$ of the inverse gauge
couplings squared. Under simplifications adopted, one has  (with
the number of generations here and in what follows $n_g=3$)
\begin{eqnarray} 
(4\pi)^2\, \beta_{g_{1}}^{(1)}&=&\frac{41}{10},\nonumber\\
(4\pi)^4\, \beta_{g_{1}}^{(2)}&=&\frac{199}{50} g_{1}^{2} -
\frac{17}{5}
y_{t}^{2} + 
\cdots. 
\end{eqnarray} 
It can be seen that at the one-loop order the coupling $g_1$ develops
a pole singularity  at $\Lambda_{g_1}^{(1)}$,
$\log\Lambda_{g_1}^{(1)}=41$. Validity of the perturbation theory in
$g_1$ restricts $\alpha_1\le4\pi$ and hence $\log\mu\le40$, which is
in the logarithmic scale
twice as large as the Planck scale. We should assume this restriction
on the physical grounds in what follows. Nevertheless, taken at its
face value the
two-loop RG has the meaning by itself. So in order to  understand
the mathematical structure of its solutions better, we extend them up
to the singularity point $\Lambda_{g_1}$.

As it is seen from Fig.~1, the actual influence of
$y_f(\lambda(\mu))$ on 
$g_1$ in two loops is somewhat sizable only for heavy Higgs.  It
diminishes the slope of $g_1(\mu)$ at $\mu$ beyond the Planck scale,
where $y_f$
are large, and shifts the singularity position
$\Lambda_{g_1}^{(2)}$ upwards to $\log\Lambda_{g_1}^{(2)}=47$ for the 
 heavy Higgs ($m_H(M_Z)=450$ GeV), which is close to that maximally
 allowed by  the perturbative consistency in $\lambda$. The value
 ~$m_H(M_Z)=136.1$ GeV
corresponds to the highest lower bound
of the electroweak vacuum stability (see later on). Curves
corresponding to the lighter Higgs bosons are very close to that for
$m_H(M_Z)=136.1$ GeV.
Curves for the intermediate values of $m_H(M_Z)$, i.e.\
$136.1$ GeV $<m_H(M_Z)<450$ GeV, are located in between the two
extreme cases. In other respects the picture in Fig.~1 is well-known.

\subsection*{{(\bf\em ii}) Yukawa  sector}

Fig.~2 depicts the evolution of the Yukawa couplings
$y_f$ for the third family SM fermions: $t$, $b$  quarks, and $\tau$
lepton, with the fact of the $t$ quark being heavy  taken
into account.  One has approximately for the top quark
\begin{eqnarray}
(4\pi)^2\, \beta_{y_{t}}^{(1)}&=& 9 y_{t}^{2} - \frac{17}{20}
g_{1}^{2} 
- \frac{9}{4} g_{2}^{2} - 8 g_{3}^{2} +\cdots ~,
\nonumber\\
(4\pi)^4\, \beta_{y_{t}}^{(2)}&=& -48 y_{t}^{4} +
\frac{3}{2}\lambda^2 - 
12 y_{t}^{2}\lambda  + \cdots ~.
\end{eqnarray}
In one loop, $\beta_{f}^{(1)}$ are dominated by the negative gauge
contributions, so that all the $y_f$ are falling down with $\mu$ and
lie in the weak coupling regime.\footnote{To be more precise, the 
one-loop trajectory for the $\tau$ lepton is mildly convex, so that
it intersects with the curve for the $b$ quark near the GUT scale.}

  But in two loops the behaviour changes
drastically. An approximate UV stable  fixed point appears at
$y_t^{(\mbox{\scriptsize UV})}\simeq 5.4$
due to compensation of the $y_{t}^{2}$ and $y_{t}^{4}$ contributions.
In the real world, this critical value is approached from below both
for $t$, $b$ quarks and for $\tau$ lepton, the faster the heavier
Higgs boson is.
Hence for the sufficiently heavy Higgs, $m_H(M_Z)\ge 200$ GeV,  all
the third family fermions would fall into the strong coupling regime
at sufficiently high $\mu$. This would make the  third family
fermions much more alike at the
high scales than at the electroweak one. In practice, prior to
$M_{Pl}$ the strong coupling develops only for $t$ quark when Higgs
is rather heavy, $m_H(M_Z)\ge 450$ GeV.
Because from the combined LEP data on the precision experiments  it
follows that $M_H\le 200$ GeV at 95\% C.L.~\cite{q98}, one may
conclude that the Yukawa sector of the
SM is weakly coupled along all the physically reasonable region of
$\mu$, $\mu\le M_{Pl}$.

\subsection*{({\bf\em iii}) Higgs  sector }

Fig.~3 presents running of the Higgs quartic coupling. One has
approximately for it
\begin{eqnarray}      
(4\pi)^2\, \beta_{\lambda}^{(1)}&=& 12 \lambda^2 -  48 y_{t}^4+
 \frac{27}{100} g_{1}^{4}+ (24 y_{t}^{2}- \frac{9}{5} g_{1}^{2})
 \lambda  + \cdots \, , \nonumber\\
(4\pi)^4\, \beta_{\lambda}^{(2)}&=& -78 \lambda^3 - 3.411 g_{1}^6 +
\cdots \, .
\end{eqnarray}
The $\beta$ function for the pure Higgs sector is known in the
$\overline{\rm MS}$ scheme up to the
three-loop order~\cite{kazakov} 
\begin{equation}      
(4\pi)^6\,\beta_{\lambda}^{(3)}=\Big(897+504\zeta(3)\Big)\lambda^4
\end{equation}
being scheme dependent.
Fig.~3a shows the one-loop behaviour of $\lambda^{1/2}$. It is seen
that for
$m_H>m_{H~min}^{(1)}=142.7$ GeV a singularity in $\lambda$ develops
at $\log\Lambda_{\lambda}^{(1)}\le 41$.
On the other hand, the theory at $m_H(M_Z)<m_{H~min}^{(1)}$ possesses
the unstable Higgs vacuum with $\lambda<0$ due to the negative
top quark contribution $\sim y_t^4$.

In two loops, there are three critical curves shown in bold in
Fig.~3b.
First of all, there appears an approximate UV stable fixed
point at $\lambda^{1/2}_{\mbox{\scriptsize UV}} \simeq 4.93$ produced
by the compensation of
the one- and two-loop terms: $\lambda^2$ and $\lambda^3$. It
corresponds
to boundary value of the Higgs mass $m_{H~max}^{(2)}(M_Z)=1200$ GeV,
at and above which the theory is definitely strongly coupled. 
The boundary Higgs mass for the vacuum instability shifts in two
loops to $m_{H~min}^{(2)}(M_Z)=136.1$ GeV. The third critical value
$m_{H~inter}^{(2)}(M_Z)=156.7$ GeV borders the region with the
potentially 
strongly coupled Higgs from the one with the weakly coupled Higgs.
Note that theory with $m_{H~min}^{(2)}<m_H(M_Z)<m_{H~inter}^{(2)}$ is
consistent in two loops up to the ultimate scale
$\mu=\Lambda_{g_1}^{(2)}$.

Finally, one can impose the requirement of the SM self-consistency
up to some cutoff scale $\Lambda$. In other terms, the theory should
be neither strongly coupled nor unstable at $\mu\le\Lambda$.
In one loop, this means that the $\lambda$ singularity
position fulfills the requirement $\Lambda_{\lambda}^{(1)} \ge
\Lambda$,
and simultaneously one has $\mu\mid_{\lambda=0} \ge \Lambda$.
In two loops, we should choose as a criterion for the onset of the
strong
coupling regime the requirements $\beta_\lambda^{(2)}/
\beta_\lambda^{(1)}|_\Lambda$ and $ \beta_\lambda^{(3)}/
\beta_\lambda^{(2)}|_\Lambda<1 $ which could guarantee the
perturbativity and the scheme independence. 
In neglect by all the couplings but $\lambda$ this would  mean that
$\lambda^{1/2}\leq 2$, in particular $m_H(M_Z)\leq 500$ GeV, the
restriction we retain for the whole SM.
Not knowing $\beta_t^{(3)}$ we restrict ourselves just by the
requirement  that $\beta_t^{(2)}/ \beta_t^{(1)}|_\Lambda\ll 1 $ which
is definitely fulfilled at $y_t\leq 2< y_t^{(UV)}$.

The one- and two-loop restrictions are drawn in Fig.~4. Here use is
made of the exact one-loop  Eq.~(\ref{delta_H}) for transition from
the $\overline{\rm MS}$
value $m_H(M_Z)$ to the physical Higgs mass $M_H$.
The sensitivity of the allowed region of the Higgs mass 
to the uncertainty of the top quark mass
is also indicated.  
Strictly speaking, allowed is the region between the most upper and 
the most lower curves. This means that for $\Lambda=M_{Pl}$ the
legitimate Higgs mass  is $M_H = (161.3 \pm 20.6)^{+4}_{-10}$ GeV.
One gets also  the lower bound  $M_H\ge 140.7\pm 10$ GeV at such a
cutoff.  The allowed Higgs mass
region is much wider for a lower cutoff scale $\Lambda$.\footnote{
The true condition for the electroweak vacuum stability is the
existence of a global minimum in the Higgs effective
potential~\cite{altarelli,casas}.
For the two-loop RG improved one-loop effective potential
$V_{eff}(\mu,\phi)$ this turns out at 
$\Lambda\simeq M_{Pl}$ to be in practice equivalent  to our
requirement that the  running coupling does not become
negative. At the lower values of $\Lambda$ there are discrepancies
which we attribute to the difference of the stability criteria.} For
completeness, we present in Fig.~5 the plot for $v(\mu)$ both in one
and two loops. It is seen
that the electroweak symmetry never restores prior to the Plank
scale.  

\section{ The fourth chiral family }

The two-loop RG global profile of the SM being understood, one
is in a position to discuss the SM conceivable extensions.
We consider here the minimum SM extension by means of the additional
heavy fermion families. If alone, the fourth family should have with
necessity the same
chirality pattern as the three light families. This is to be required
to avoid the potential problem of  the large direct mass mixing
for the fourth family with the light ones.

  What concerns the fifth family, there are two possibilities: either
it has the same chirality as the four previous families, or it is a
mirror one
(or to state it differently, it is charge conjugate with respect to
the rest of the families). In the first case, the analysis repeats
itself just with more parameters. In the second case, the large
direct mass terms could be
introduced for the pair of the fourth and fifth families, in addition
to Yukawa couplings. This proliferates enormously the number of free
parameters and makes the general analysis impossible. On the other
hand, if one chooses a mass independent renormalization scheme, say 
$\overline{\rm MS}$, the net influence of the direct mass terms on
the evolution of the SM parameters will be just in the threshold
effects. Barring them, this case, which may likewise be attributed to
one vector-like family, is technically equivalent to the case with
two chiral families.

 For these reasons, we restrict ourselves by one new chiral family.
In order to conform with experimental value for the number 
of light neutrinos ($n_{\nu}=3$), we should add also the right-handed 
neutrinos $\nu_R$ (at least for the fourth family) and the proper
Yukawa couplings for them. The right-handed neutrinos may possess the
explicit Majorana mass as well, so that the physical neutrino masses
may be quite different from their Yukawa counterparts. Because in the
mass independent
renormalization the explicit mass terms are important only in the
threshold effects, we disregard them in what follows. We generalized
the two-loop SM $\beta$ functions of Ref.~\cite{machacek} to the case
with the neutrino Yukawa couplings. The results are given in the
Appendix A2.
The rest of the $\beta$ functions is as in Ref.~\cite{machacek}.
For practical calculations with the fourth family we neglected by
the light neutrino Yukawa couplings.
The reduced four family $\beta$ functions are given in Appendix A3.

At present, there are no theoretical hints on
the existence (or v.v.) of the fourth (and the higher) family.
Nevertheless, one can extract some restrictions on the corresponding
fermion masses. They are twofold, the direct and indirect ones, being
in a sense complementary to each other. The first group gives bounds
on the common mass scale of the fourth family, the second one
restricts the splitting of the masses inside the family.

The existing direct experimental bounds on the masses of the fourth
fami\-ly quarks $t_4$ and $b_4$ depend  somewhat on the assumptions
about their decays. If the lightest of the quarks, say $b_4$, is
stable enough to leave the 
detector, the limit on its mass is $M_4 \ge 140$ GeV \cite{cdf}.
On the other hand, for unstable quarks, decaying inside the detector,
the limit can be estimated from the CDF and D0 searches for the top
quark~\cite{top} to be  about $M_t$. What concerns the neutral and
charged  leptons of the fourth family, $\nu_4$  and $e_4$, it follows
from LEP searches that $M_{\nu_4} \ge 59$ GeV and $M_{e_4} \ge 90$
GeV  at 95\% C.L.~\cite{rpp,lep}.

The indirect restrictions can be extracted from the precision
electroweak data, and they are related to the absence of decoupling
with respect to the heavy chiral fermions. This results in the
quadratically growing dependence of
the electroweak radiative corrections on the heavy fermion
masses. To avoid such a large  corrections,
as the precision data require, 
the members of a heavy fermion doublet should be highly degenerate.
Namely, one should have for the quarks $t_4$ and $b_4$ that 
$(M_{t_4}^2 - M_{b_4}^2)/M_Z^2 \le 1$, and similarly for the leptons
$\nu_4$, $e_4$.\footnote{One important peculiarity of the
vector-like family is the decoupling with respect to the explicit
mass term, at the Yukawa couplings being fixed. Hence, unlike the
chiral family case, there is no need here for the high degeneracy in
the  Yukawa couplings to suppress the large radiative corrections.} 

To reduce the number of free parameters we assume in what follows
that $m_{t_4} = m_{b_4} = m_Q$ and $m_{\nu_4} = m_{e_4} = m_L$.
As representative, we considered two cases: $m_L/m_Q = 1/2$ and 1, 
with the common mass $m_Q$ of the heavy quarks given by the fourth
family scale $m_4$. It follows that both these typical cases do not
contradict to the direct
experimental bounds if $m_4 \ge 180$ GeV. Our results for the case
$m_4 = 200$ GeV, which we consider 
as more realistic, are presented in Figs.~6--9. Cases~$a$ correspond
to $m_L/m_Q = 1/2$ and cases~$b$ to $m_L/m_Q = 1$.

Fig.~6 shows the evolution of $\alpha_i^{-1}$ with 
$\mu$.\footnote{It may be noted that the GUT triangle
shrinks  but the conceivable gauge unification takes place
beyond the region of perturbativity both in Yukawa and Higgs
sectors.}  
It is seen that the two-loop contributions manifest themselves at
rather low scales, $\mu = (10^7 - 10^8)$ GeV. They are governed by
the onset of the strong coupling regime in the Yukawa sector at such
a $\mu$ (see Fig.~7).
Accordingly, the perturbatively consistent region of $\mu$ in the
Higgs sector shrinks to the same values (see Fig.~8). Applying now
the same criteria of self-consistency as in the case of the minimal
SM we get the allowed values
of $M_H$ depending on the cutoff scale $\Lambda$ (Fig.~9). 
The sensitivity to the shift in the mass $m_4$ is also indicated.
The dependence on $\Delta M_t$ is much smaller, and it is not shown.

Finally, Figs.~10--11 present the one- and two-loop allowed
regions in the $m_4$--$M_H$ plane.  The influence of the Yukawa
perturbative validity in two loops on the allowed regions of $m_4$
and $M_H$  is rather weak at the high $\Lambda$. 
Fig.~11 excludes the fourth heavy chiral family at high 
$\Lambda$, $\Lambda \ge 10^{10}$~GeV, independent of the Higgs mass.
Under LEP restriction $M_H\leq 200$ GeV, the fourth chiral family is
completely excluded.\footnote{
Though one possible caveat emerges if one adopts that the fourth
family is vector-like and that, unnatural as it may seem, its Yukawa
couplings are small. Then the ensuing restrictions on the  family are
strongly reduced, and the vector-like fourth family could exist.}
Dependence of the restrictions on the top mass uncertainty is very
faint.

\section { Conclusions }

Let us summarize the differences in the RG global profiles of the
SM with three and four chiral generations. For three generations with 
the experimentally known masses,
the Yukawa sector is weakly coupled in the one-loop approximation.
Prior to the Planck scale,
the strong coupling may appear in one loop only
in the Higgs self-interactions for the sufficiently heavy Higgs.
It drives strong coupling for the Yukawa sector as well, but only
through two loops. As a result, this influence is reduced, and the
Yukawa sector stays weakly coupled up to the Planck scale for all
experimentally preferred values of the Higgs mass, $M_H \le 200$ GeV.
Validity of the perturbative SM up to the Planck scale,
the Yukawa sector including, as well as the vacuum stability require
the Higgs mass to be $M_H = (161.3 \pm 20.6)^{+4}_{-10}$ GeV  and
$M_H\ge 140.7^{+10}_{-10}$ GeV.  Here  the $M_H$ corridor is  the
theoretical one with the  errors being produced by  the top mass
uncertainty. The allowed Higgs mass region is wider for a lower
cutoff scale~$\Lambda$.

The inclusion of the fourth heavy chiral family qualitatively changes
the mode of the SM realization.
With the addition of the family, the strong coupling is driven 
in one loop by the Yukawa interactions. It transmits to the Higgs
self-interactions at the one-loop order, too. 
Hence the strong coupling develops in both these sectors
in parallel, and their couplings blow up at sufficiently low scales. 
As a result, the requirement of self-consistency of the perturbative
SM as an underlying theory up to the Planck or GUT scale excludes
the fourth chiral family. But as an effective theory, the SM allows
the heavy chiral family with the mass up to 250 GeV depending on the
Higgs mass and the cutoff scale.  Under precision experiment
restriction $M_H\leq 200$~GeV, the fourth chiral family,  taken
alone, is excluded. Nevertheless a pair of the chiral families
constituting  the vector-like one  could still exist.

\paragraph{Acknowledgement}
This work was supported by the RFBR under grant No.~96--02--18122.

\paragraph{Note added in the 2nd version of the e-print}
The authors wish to thank Akin Wingerter for the discussion resulted in 
a correction of misprints in contributions to $\beta^{(2)}_{{\bf Y}_e}$
induced by heavy neutrino (pp. 18, 21). 

\paragraph{Note added in the 3rd version of the e-print}
The update includes correction of a typo in the 
$-9 \sum_g (3 y_{u_g}^4 + 3 y_{d_g}^4  - \dots)$ 
term in the expression for $\beta^{(2)}_{y_\tau}$ in Appendix A1,
and footnotes clarifying the expressions for heavy neutrino contributions to 
$\beta$-functions for Yukawa couplings in Appendices A2, A3.
The authors are grateful to Akin Wingerter for pointing to these issues 
in Ref.~\cite{Wingerter2011}.

\newpage

\section*{ Appendices }

\subsection*{A1\quad SM $\beta$ functions}

Here $n_g = 3$ and the generation index $g$ runs over the values $g =
1,2,3$. One has in fact  $\sum_g \equiv \delta_{g3}$. Here and in
what follows  the parts of the expressions for the anomalous
dimension $\gamma_v$ which are proportional to the gauge couplings
are valid in the 't~Hooft--Landau gauge~$\xi = 0$. 

\def\ys{\sum_g (3 y_{u_g}^2 + 3 y_{d_g}^2 +  y_{e_g}^2)}

\def\h4s{9 \sum_g (3 y_{u_g}^4 + 3 y_{d_g}^4 - \frac{2}{3} y_{u_g}^2
y_{d_g}^2
+ y_{e_g}^4)}

\def\y4s{(\frac{17}{20} g_1^2 + \frac{9}{4} g_2^2 + 8 g_3^2) \sum_g
y_{u_g}^2 
\\ + (\frac{1}{4} g_1^2 + \frac{9}{4} g_2^2 + 8 g_3^2) \sum_g
y_{d_g}^2 
\\ + (\frac{3}{4} g_1^2 + \frac{3}{4} g_2^2) \sum_g y_{e_g}^2} 

\def\Tr{\mbox{Tr}}

\subsubsection*{ One-loop contributions }

({\em i}) Gauge sector: 

\vspace*{2ex}

\begin{tabular}{rcl}
$(4\pi)^2 g_1^{-3}\beta^{(1)}_{g_1}$ &=& $\frac{41}{10}$,\\[1 ex]    
$(4\pi)^2g_2^{-3}\beta^{(1)}_{g_2}$ &=& $- \frac{19}{6}$,\\[1 ex]
$(4\pi)^2g_3^{-3}\beta^{(1)}_{g_3}$ &=& $- 7$.
\end{tabular}

\vspace*{2ex}

\noindent
({\em ii}) Yukawa sector: 

\vspace*{2ex}

\begin{tabular}{rcl}
$(4\pi)^2 y_\tau^{-1}\beta^{(1)}_{y_\tau}$ &=& 
$3 y_\tau^2 + 2 \ys - \frac{9}{4} g_1^2 -
\frac{9}{4} g_2^2$,\\[1ex]     

$(4\pi)^2 y_t^{-1}\beta^{(1)}_{y_t}$ &=& 
$ 3 y_t^2 - 3 y_b^2 + 2 \ys - \frac{17}{20} g_1^2 - \frac{9}{4} g_2^2
- 8 g_3^2$,\\[1 ex]     

$(4\pi)^2 y_b^{-1}\beta^{(1)}_{y_b}$ &=& 
$ 3 y_b^2 - 3 y_t^2 + 2 \ys - \frac{1}{4} g_1^2 - \frac{9}{4} g_2^2
- 8 g_3^2$.
\end{tabular}

\vspace*{2ex}

\noindent
({\em iii}) Higgs sector: 

\vspace*{2ex}

\begin{tabular}{rcl}
$(4\pi)^2\beta^{(1)}_{\lambda}$ &=&
$12 \lambda^2 + 8 \lambda \ys - 9 \lambda (\frac{1}{5}g_1^2 + 
g_2^2)$\\[1 ex]
&&$- 16 \sum_g (3 y_{u_g}^4 + 3 y_{d_g}^4 + y_{e_g}^4)
+ \frac{9}{4} (\frac{3}{25} g_1^4 +  g_2^4 + \frac{2}{5} g_1^2
g_2^2)$,\\[1 ex]

$(4\pi)^2v^{-1}\gamma^{(1)}_{v}$ &=&
$ -2 \ys + \frac{9}{4} (\frac{1}{5} g_1^2 +  g_2^2)$.\\
&&\\
\end{tabular} 

\subsubsection*{ Two-loop contributions }

({\em i}) Gauge sector: 

\vspace*{2ex}

\begin{tabular}{rcl}
$(4\pi)^4g_1^{-3}\beta^{(2)}_{g_1}$ &=& 
$  \frac{199}{50} g_1^2 + \frac{27}{10} g_2^2 + \frac{44}{5} g_3^2   
- \sum_g (\frac{17}{5} y_{u_g}^2 + y_{d_g}^2 
+ 3 y_{e_g}^2)$,\\[1 ex] 

 $(4\pi)^4g_2^{-3}\beta^{(2)}_{g_2}$ &=&
$   \frac{9}{10} g_1^2 + \frac{35}{6} g_2^2 +
12 g_3^2   
  - \sum_g (3 y_{u_g}^2 + 3 y_{d_g}^2 + y_{e_g}^2)$,\\[1 ex] 

$(4\pi)^4g_3^{-3}\beta^{(2)}_{g_3}$ &=&
$   \frac{11}{10} g_1^2 + \frac{9}{2} g_2^2 - 26 g_3^2   
  - 4 \sum_g (y_{u_g}^2 + y_{d_g}^2)$.
\end{tabular}

\vspace*{2ex}

\noindent
({\em ii}) Yukawa sector: 

\vspace*{2ex}

\begin{tabular}{rcl}
$(4\pi)^4 y_\tau^{-1} \beta^{(2)}_{y_\tau}$ &=&
$ 6 y_\tau^4 - 9 y_\tau^2 \ys - 9 \sum_g (3 y_{u_g}^4 + 3
y_{d_g}^4$\\[1 ex] 
&& $ - \frac{2}{3} y_{u_g}^2 y_{d_g}^2 + y_{e_g}^4)
     + \frac{3}{2} \lambda^2 - 12 \lambda y_\tau^2 
     + (\frac{387}{40} g_1^2 + \frac{135}{8} g_2^2)y_\tau^2$\\[1 ex] 
&& $ + 5(\frac{17}{20} g_1^2 + \frac{9}{4} g_2^2 + 8
g_3^2)
\sum_g y_{u_g}^2 + 5(\frac{1}{4} g_1^2 + \frac{9}{4} g_2^2 + 8
g_3^2)
\sum_g y_{d_g}^2$
\\[1 ex] 
&& $ +\frac{15}{4} ( g_1^2 +  g_2^2) \sum_g
y_{e_g}^2 
     + \frac{1371}{200} g_1^4 + \frac{27}{20} g_1^2 g_2^2 -
     \frac{23}{4} g_2^4$,\\
&&\\

$(4\pi)^4 y_t^{-1} \beta^{(2)}_{y_t}$ &=&
$ 6 y_t^4 - 5 y_t^2 y_b^2 + 11 y_b^4 + (5 y_b^2 - 9 y_t^2) \ys $\\[1
ex] 
&& $ - \h4s 
     + \frac{3}{2} \lambda^2 - 4 \lambda(3 y_t^2 + y_b^2)$\\[1 ex] 
&& $ + (\frac{223}{40} g_1^2 + \frac{135}{8} g_2^2 + 32 g_3^2) y_t^2 
     - (\frac{43}{40} g_1^2 - \frac{9}{8} g_2^2 + 32
     g_3^2)y_b^2$\\[1 ex] 
&& $ + 5(\frac{17}{20} g_1^2 + \frac{9}{4} g_2^2 + 8
g_3^2) 
\sum_g y_{u_g}^2
 + 5(\frac{1}{4} g_1^2 + \frac{9}{4} g_2^2 + 8 g_3^2)
 \sum_g  y_{d_g}^2$
\\[1 ex] 
&& $ + \frac{15}{4}( g_1^2 +  g_2^2) \sum_g
y_{e_g}^2$
\\[1 ex] 
&& $ + \frac{1187}{600} g_1^4 - \frac{9}{20} g_1^2 g_2^2 
 - \frac{23}{4} g_2^4 + \frac{19}{15} g_1^2g_3^2  + 9 g_2^2 g_3^2 -
 108
g_3^4$,\\
&&\\

$(4\pi)^4 y_b^{-1} \beta^{(2)}_{y_b}$ &=&
$ 6 y_b^4 - 5 y_b^2 y_t^2 + 11 y_t^4 + (5 y_t^2 - 9 y_b^2) \ys 
$\\[1ex] 
&& $ - \h4s  
     + \frac{3}{2} \lambda^2 - 4 \lambda(3 y_b^2 + y_t^2)$\\[1 ex] 
&& $ + (\frac{187}{40} g_1^2 + \frac{135}{8} g_2^2 + 32 g_3^2)y_b^2 
     - (\frac{79}{40} g_1^2 - \frac{9}{8} g_2^2 + 32
     g_3^2)y_t^2$\\[1 ex] 
&& $ + 5(\frac{17}{20} g_1^2 + \frac{9}{4} g_2^2 + 8
g_3^2) 
\sum_g y_{u_g}^2 
     + 5(\frac{1}{4} g_1^2 + \frac{9}{4} g_2^2 + 8 g_3^2) \sum_g
     y_{d_g}^2$
\\[1 ex] 
&& $ + \frac{15}{4}( g_1^2 +  g_2^2) \sum_g
y_{e_g}^2$
\\[1 ex] 
&& $ - \frac{127}{600} g_1^4 - \frac{27}{20} g_1^2 g_2^2
  - \frac{23}{4} g_2^4   + \frac{31}{15} g_1^2 g_3^2  + 9 g_2^2 g_3^2
     - 108 g_3^4$.
\end{tabular}  


\newpage

\noindent
({\em iii}) Higgs sector: 

\vspace*{2ex}

\begin{tabular}{rcl}
$(4\pi)^4\beta^{(2)}_{\lambda}$
&=&$ - 78 \lambda^3 + 54 \lambda^2 (\frac{1}{5} g_1^2 +  g_2^2)
     + \lambda (\frac{1887}{200} g_1^4 + \frac{117}{20} g_1^2 g_2^2
                - \frac{73}{8} g_2^4) $\\[1 ex]
&& $ - \frac{3411}{1000} g_1^6 
     - \frac{1677}{200}  g_1^4 g_2^2 
     - \frac{289}{40} g_1^2 g_2^4 + \frac{305}{8} g_2^6 $\\[1 ex]	 
&& $ - 3 g_2^4 \ys $\\[1 ex]
&& $ - \frac{32}{5} g_1^2 \sum_g (2 y_{u_g}^4 - y_{d_g}^4 + 3
y_{e_g}^4) - 256 g_3^2 \sum_g (y_{u_g}^4 + y_{d_g}^4)
      $\\[1 ex]
&& $ + 20 \lambda \mbox{\Big (}(\frac{17}{20} g_1^2 + \frac{9}{4}
g_2^2 +
     8 g_3^2) \sum_g y_{u_g}^2
     + (\frac{1}{4} g_1^2 + \frac{9}{4} g_2^2 + 8 g_3^2) \sum_g
     y_{d_g}^2$
\\[1 ex] 
&& $ + \frac{3}{4}( g_1^2 +  g_2^2) \sum_g
y_{e_g}^2\mbox{\Big )} 

     + \frac{6}{5} g_1^2 \mbox{\Big (} (-\frac{57}{10} g_1^2 +
     21
     g_2^2) \sum_g y_{u_g}^2$
\\[1 ex] 
&& $ + 3(\frac{1}{2} g_1^2 + 3  g_2^2) \sum_g y_{d_g}^2  
     +  (-\frac{15}{2} g_1^2 + 11 g_2^2) \sum_g y_{e_g}^2
     \mbox{\Big
     )} $
\\[1 ex]
&& $ - 48 \lambda^2 \ys $\\[1 ex] 
&& $ - 4 \lambda \sum_g (3 y_{u_g}^4 + 3 y_{d_g}^4 + y_{e_g}^4
     - 6 y_{u_g}^2 y_{d_g}^2) $\\
&& $ + 160 \sum_g (3 y_{u_g}^6 + 3 y_{d_g}^6 + y_{e_g}^6)
     - 96 \sum_g (y_{u_g}^4 y_{d_g}^2 + y_{d_g}^4 y_{u_g}^2)
     $,\\[1 ex]	 
$(4\pi)^4v^{-1}\gamma^{(2)}_{v}$
&=&$-\frac{3}{2}\lambda^2 
- \frac{1293}{800}g_1^4 + \frac{271}{32}g_2^4 -
\frac{27}{80}g_1^2g_2^2$\\[1ex]
&&$-\frac{5}{2}(\frac{17}{10}g_1^2 +\frac{9}{2}g_2^2 +16 g_3^2)
\sum_g y_{u_g}^2
-\frac{5}{2}(\frac{1}{2}g_1^2 +\frac{9}{2}g_2^2 +16 g_3^2)
\sum_g y_{d_g}^2$\\[1ex]
&&$-\frac{15}{4}(g_1^2 +g_2^2) \sum_g y_{e_g}^2 
+ 9\sum_g (3 y_u^4 +3 y_d^4  -\frac{2}{3} y_u^2
y_d^2 +  y_e^4)  $.\\
\end{tabular}

\subsection*{A2\quad Neutrino Yukawa contributions to SM $\beta$ 
functions}

\def\y{\mbox{\bf Y}}
\def\yn2{\mbox{\bf Y}_{\nu}^\dagger \mbox{\bf Y}_{\nu}}
\def\ye2{\mbox{\bf Y}_{e}^\dagger \mbox{\bf Y}_{e}}
\def\yu2{\mbox{\bf Y}_{u}^\dagger \mbox{\bf Y}_{u}}
\def\yd2{\mbox{\bf Y}_{d}^\dagger \mbox{\bf Y}_{d}}

\def\ynu{ \mbox{\scriptsize\bf Y}_\nu }
\def\yee{ \mbox{\scriptsize\bf Y}_e }
\def\yuu{ \mbox{\scriptsize\bf Y}_u }
\def\ydd{ \mbox{\scriptsize\bf Y}_ d }

\subsubsection*{ One-loop contributions}

({\em i}) Yukawa sector:\footnote{%
	The expression for $\beta^{(1)}_{\ynu}$ is complete.
	For economy of space,
	in $\Delta\beta^{(1)}_{\yee, \yuu, \ydd}$ only neutrino induced contributions are shown explicitly.
	The full expressions can be recovered using Ref.~\cite{machacek}.
   }

\vspace*{2ex}

\begin{tabular}{rcl}
$ (4\pi)^2 \y_{\nu}^{-1} \beta^{(1)}_{\ynu} $ 
&=&
$ \frac{3}{2} (\yn2 - \ye2)
  + Y_2(S) - \frac{9}{20} g_1^2 - \frac{9}{4} g_2^2 $,\\[1 ex]    
$ (4\pi)^2 \y_e^{-1}\Delta\beta^{(1)}_{\yee} $ &=&
$  - \frac{3}{2} \yn2 + \Tr (\yn2)$,\\[1 ex] 
$ (4\pi)^2 \y_u^{-1}\Delta\beta^{(1)}_{\yuu}$ &=& $ \Tr
(\yn2)$,\\[1 ex]     
$ (4\pi)^2 \y_d^{-1}\Delta\beta^{(1)}_{\ydd}$ &=& $ \Tr
(\yn2)$.
\end{tabular}

\vspace*{2ex}

\noindent
({\em ii}) Higgs sector: 

\vspace*{2ex}

\begin{tabular}{rcl}
$(4\pi)^2\Delta\beta^{(1)}_{\lambda}$ &=&
$ 4\lambda \Tr (\yn2) - 4\Tr \mbox{\Big (}(\yn2)^2\mbox{\Big )}$,
\nonumber\\ [1 ex]
$(4\pi)^2v^{-1}\Delta\gamma^{(1)}_{v}$ &=&
$ - \Tr (\yn2)$.
\end{tabular}

\subsubsection*{ Two-loop contributions }

({\em i}) Gauge sector: 

\vspace*{2ex}

\begin{tabular}{r c l}
$(4\pi)^4g_1^{-3}\Delta\beta^{(2)}_{g_1}$ &=& $- \frac{3}{10} \Tr
(\yn2)$,\\[1 ex]
$(4\pi)^4g_2^{-3}\Delta\beta^{(2)}_{g_2}$ &=& $- \frac{1}{2} \Tr
(\yn2)$.
\end{tabular}

\vspace*{2ex}

\noindent
({\em ii}) Yukawa sector:\footnote{%
	The expression for $\beta^{(2)}_{\ynu}$ is complete.
	In $\Delta\beta^{(2)}_{\yee}$ neutrino induced contributions 
	included into $Y_4(S)$ and $\chi_4(S)$ invariants are not shown explicitly
	for economy of space in this Appendix, 
	though are fully taken into account in the calculations in the main text.
    $\Delta\beta^{(2)}_{\yuu, \ydd}$ include all neutrino induced terms.
	The full expressions can be recovered using Ref.~\cite{machacek}.
  }

\vspace*{2ex}

\begin{tabular}{r c l}
$(4\pi)^4 \y_\nu^{-1} \beta_{\ynu}^{(2)}$ &=&
$\frac{3}{2}(\yn2)^2  - \yn2\ye2 - \frac{1}{4} \ye2\yn2
+ \frac{11}{4} (\ye2)^2$\\[1 ex] 
&& $ + Y_2(S)(\frac{5}{4} \ye2 - \frac{9}{4} \yn2) - \chi_4(S) $\\[1
ex]
&& $ + \frac{3}{2} \lambda^2 - 2 \lambda (3\yn2 + 
\ye2)$\\[1 ex]    
&& $ + (\frac{279}{80} g_1^2 + \frac{135}{16} g_2^2) \yn2
- (\frac{243}{80} g_1^2 - \frac{9}{16} g_2^2) \ye2
+ \frac{5}{2} Y_4(S)$\\[1 ex]     
&& $+ ( - \frac{3}{40} + \frac {1}{5} n_g) g_1^4 - \frac{27}{20}
g_1^2 g_2^2
- (\frac{35}{4} - n_g) g_2^4$,\\[1 ex]
$(4\pi)^4 \y_e^{-1} \Delta\beta^{(2)}_{\yee}$ &=&
$  - \ye2\yn2 - \frac{1}{4} \yn2\ye2
+ \frac{11}{4} (\yn2)^2 $
\\[1 ex]    
&&$ + \frac{5}{4} Y_2(S)  \yn2
- 2 \lambda \yn2 + (- \frac{27}{16} g_1^2 + \frac{9}{16}
g_2^2) \yn2 $,\\[1ex]
$(4\pi)^4 \y_u^{-1} \Delta\beta^{(2)}_{\yuu}$ &=&
$ (\frac{5}{4} \yd2 - \frac{9}{4} \yu2) \Tr (\yn2)
- \Tr \mbox{\Big (}\frac{9}{4} (\yn2)^2 $
\\[1 ex]
&&$ -\frac{1}{2}\yn2\ye2\mbox{\Big )}
+ \frac{15}{8} (\frac {1}{5} g_1^2 +  g_2^2) \Tr (\yn2)$,\\[1ex]
$(4\pi)^4 \y_d^{-1} \Delta\beta^{(2)}_{\ydd}$ &=&
$ (\frac{5}{4} \yu2 - \frac{9}{4} \yd2) \Tr (\yn2)
- \Tr \mbox{\Big (} \frac{9}{4} (\yn2)^2 $
\\[1 ex]
&&$- \frac{1}{2} \yn2\ye2\mbox{\Big )}
+ \frac{15}{8} (\frac{1}{5} g_1^2 +  g_2^2) \Tr (\yn2)$.
\end{tabular}


\newpage

\noindent
({\em iii}) Higgs sector: 

\vspace*{2ex}

\begin{tabular}{r c l}
$(4\pi)^4\Delta\beta^{(2)}_{\lambda}$ &=&
$-\frac{3}{2} g_2^4 \Tr (\yn2) 
- \frac{3}{10} g_1^2 (\frac{3}{5}g_1^2 +2 g_2^2) \Tr (\yn2)$\\[1
ex]
&& $+ \frac{15}{2}\lambda (\frac{1}{5} g_1^2+ g_2^2) \Tr (\yn2) -
24\lambda^2 \Tr (\yn2)$\\[1 ex]
&& $- \lambda \Tr \mbox{\Big( }(\yn2)^2\mbox{\Big )}
 + 2\lambda \Tr \mbox{\big (}\yn2\ye2\mbox{\big )}$\\[1 ex]
&& $+ 20 \Tr \mbox{\Big (}(\yn2)^3\mbox{\Big )}$\\[1 ex]
&& $ - 4 \Tr \mbox{\Big (}\yn2(\yn2 + \ye2)\ye2\mbox{\Big )}$,
\\[1 ex]
$(4\pi)^2v^{-1}\Delta\gamma^{(2)}_{v}$ &=&
$ -\frac{15}{8}\Tr\Big((\frac{1}{5} g_1^2+ g_2^2)\yn2\Big)$\\[1 ex]
&& $+\frac{9}{4} \Tr\Big({(\yn2)}^2\Big)
- \frac{1}{2} \Tr\big(\yn2\ye2 \big)   $\, ,
\end{tabular}

\noindent
where

\vspace*{2ex}

\begin{tabular}{r c l}
$Y_2(S)$ &=& $\Tr (3 \yu2 + 3 \yd2 + \yn2 + \ye2)$,\\[1ex]
$\chi_4(S)$ &=& $\frac{9}{4}\Tr \mbox{\Big( }3 (\yu2)^2 + 3 (\yd2)^2 
+ (\yn2)^2 $\\[1 ex] 
&&$+ (\ye2)^2- \frac{2}{3} \yu2\yd2 - \frac{2}{9} \yn2\ye2\mbox{\Big
)}$,\\[1ex]
$Y_4(S)$ &=& $\Tr \mbox{\Big (} 
(\frac{17}{20} g_1^2 + \frac{9}{4} g_2^2 + 8 g_3^2) \yu2 
$ \\[1ex]
&&$+ (\frac{1}{4} g_1^2+ \frac{9}{4} g_2^2 + 8 g_3^2) \yd2$\\[1 ex]
&& $ +\frac{3}{4} (\frac{1}{5} g_1^2 +  g_2^2) \yn2 
+\frac{3}{4}( g_1^2 +  g_2^2)\ye2 \mbox{\Big )}$.\\[1ex]
&&\\
\end{tabular}

\noindent
Our definition of invariants  generalizes immediately that of
Ref.~\cite{machacek}.

\newpage

\subsection*{ A3\quad Heavy neutrino contributions to SM $\beta$
functions }

We put here for simplicity  $\nu_4 = N, e_4 = E$.
In what follows $n_g = 4$ and the 
generation index $g$ runs over the values $g = 1,\ldots,4$.
Our normalization for the Yukawa couplings $y_{f_g}$ corresponds to
$(\mbox{\bf Y}_f^{diag})_{gg'} = \sqrt{2}\, y_{f_g} \delta_{gg'}$.
In practice, one has $y_{\nu_g}=0$ for $g\neq 4$.

\subsubsection*{ One-loop contributions}


({\em i}) Yukawa sector:\footnote{%
	The expression for $\beta^{(1)}_{y_N}$ is complete.
	For economy of space,
	in $\Delta\beta^{(1)}_{y_{{}_E}, y_{{}_f}}$ only neutrino induced contributions are shown explicitly
	(cf. Appendix~A2).
   }

\vspace*{2ex}

\begin{tabular}{r c l}
$(4\pi)^2 y_N^{-1}\beta^{(1)}_{y_{{}_N}}$ &=& $3 y_N^2 - 3 y_E^2 +
Y_2(S) 
- \frac{9}{20} g_1^2 - \frac{9}{4} g_2^2$,\\[1 ex]    
$(4\pi)^2 y_E^{-1}\Delta\beta^{(1)}_{y_E}$ &=& $- y_N^2$,\\[1 ex]     
$(4\pi)^2 y_f^{-1}\Delta\beta^{(1)}_{y_f}$ &=& $2 y_{N}^2$\,,
\end{tabular}

\vspace*{2ex}
\noindent
where $f \ne N, E$.

\vspace*{2ex}

\noindent
({\em ii}) Higgs sector: 

\vspace*{2ex}

\begin{tabular}{r c l}
$(4\pi)^2\Delta\beta^{(1)}_{\lambda}$ &=& $8\lambda y_{N}^2 - 16
y_{N}^4$.\\[1ex] 
$(4\pi)^2v^{-1}\Delta\gamma^{(1)}_{v}$ &=&
$ -2 y^2_N $.
\end{tabular}

\subsubsection*{Two-loop contributions}


({\em i}) Gauge sector: 

\vspace*{2ex}

\begin{tabular}{r c l}
$(4\pi)^4g_1^{-3}\Delta\beta^{(2)}_{g_1}$ &=& $- \frac{3}{5}
y_N^2$,\\[1 ex]
$(4\pi)^4g_2^{-3}\Delta\beta^{(2)}_{g_2}$ &=& $- y_N^2$.
\end{tabular}

\vspace*{2ex}

\noindent
({\em ii}) Yukawa sector:\footnote{%
	The expression for $\beta^{(2)}_{y_{{}_N}}$ is complete.
	In $\Delta\beta^{(2)}_{y_{{}_E}}$ neutrino induced contributions 
	included into $Y_4(S)$ and $\chi_4(S)$ invariants are not shown explicitly
	for economy of space in this Appendix, 
	though are fully taken into account in the calculations in the main text.
    $\Delta\beta^{(2)}_{y_{{}_u}, y_{{}_d}}$ include all neutrino induced terms
	(cf. Appendix~A2).
  }

\vspace*{2ex}

\begin{tabular}{r c l}
$(4\pi)^4 y_N^{-1} \beta_{y_{{}_N}}^{(2)}$ &=&
$6 y_{N}^4 - 5 y_{E}^2 y_{N}^2 + 11 y_{E}^4 
+ \frac{1}{2} Y_2(S)(5 y_{E}^2 - 9 y_{N}^2)$\\[1 ex]    
&& $- \chi_4(S) + \frac{3}{2} \lambda^2
- 4 \lambda (3y_{N}^2 +  y_{E}^2)$\\[1 ex]    
&& $+ (\frac{279}{40} g_1^2 + \frac{135}{8} g_2^2) y_N^2 
- (\frac{243}{40} g_1^2 - \frac{9}{8} g_2^2) y_E^2 
+ \frac{5}{2} Y_4(S)$\\[1 ex]     
&& $+ \frac{3}{5} ( - \frac{1}{8} + \frac {1}{3} n_g) g_1^4 
    - \frac{27}{20} g_1^2 g_2^2
- (\frac{35}{4} - n_g) g_2^4$,\\[1 ex]

$(4\pi)^4 y_E^{-1} \Delta\beta^{(2)}_{y_E}$ &=&
$ 11 y_N^4 - 5 y_E^2 y_N^2 + \frac{5}{2} Y_{2}(S) y_{N}^2$
\\[1 ex]    
&&$ - 4 \lambda y_{N}^2
- (\frac{27}{8} g_1^2 - \frac{9}{8} g_2^2) y_N^2$,\\[1 ex]

$(4\pi)^4 y_u^{-1} \Delta\beta^{(2)}_{y_u}$ &=&
$ (5 y_{d}^2 - 9 y_{u}^2) y_N^2
- (9 y_{N}^2 - 2 y_{E}^2) y_N^2
+ \frac{15}{4} (\frac{1}{5}g_1^2 +  g_2^2) y_N^2$,\\[1 ex]

$(4\pi)^4 y_d^{-1} \Delta\beta^{(2)}_{y_d}$ &=&
$(5 y_{u}^2 - 9 y_{d}^2) y_N^2
-(9 y_{N}^2 - 2 y_{E}^2) y_N^2
+ \frac{15}{4} (\frac{1}{5}g_1^2 +  g_2^2) y_N^2$.
\end{tabular}

\vspace*{2ex}

\noindent
({\em iii}) Higgs sector: 

\vspace*{2ex}

\begin{tabular}{r c l}
$(4\pi)^4\Delta\beta^{(2)}_{\lambda}$ &=& $-3 g_2^4 y_{N}^2
- \frac{3}{5} g_1^2 (\frac{3}{5} g_1^2 + 2g_2^2) y_{N}^2$\\[1ex]
&&$+ 15\lambda (\frac{1}{5}g_1^2+ g_2^2) y_{N}^2 - 48\lambda^2
y_{N}^2$\\[1 ex]
&& $- 4\lambda y_{N}^4 + 8\lambda y_{E}^2 y_{N}^2
+ 160 y_{N}^6 - 32 y_{E}^2 (y_{N}^2 + y_{E}^2) y_N^2$,\\[1 ex]
$(4\pi)^2v^{-1}\Delta\gamma^{(2)}_{v}$ &=&
$ -\frac{15}{4}(\frac{1}{5} g_1^2+ g_2^2)y_{N}^2
+9  y_{N}^4 - 2y_{N}^2 y_{E}^2 $ ,
\end{tabular}

\vspace*{2ex}

\noindent
where 

\vspace*{2ex}

\begin{tabular}{r c l}
$Y_2(S)$ &=& $2 \sum_g (3 y_{u_g}^2 + 3 y_{d_g}^2 + y_{\nu_g}^2 +
y_{e_g}^2)$,\\[1 ex]

$\chi_4(S)$ &=& $9 \sum_g (3 y_{u_g}^4 + 3 y_{d_g}^4
- \frac{2}{3} y_{u_g}^2 y_{d_g}^2                         
+ y_{\nu_g}^4 + y_{e_g}^4 - \frac{2}{9} y_{\nu_g}^2 y_{e_g}^2
)$,\\[1 ex]

$Y_4(S)$ &=& $2 \sum_g \mbox{\Big (} 
(\frac{17}{20} g_1^2 + \frac{9}{4} g_2^2 + 8 g_3^2) y_{u_g}^2
+(\frac{1}{4} g_1^2 + \frac{9}{4} g_2^2 + 8 g_3^2) y_{d_g}^2$\\[1
ex]
&& $+\frac{3}{4}(\frac{1}{5} g_1^2 +  g_2^2) y_{\nu_g}^2 
+\frac{3}{4}( g_1^2 +  g_2^2) y_{e_g}^2 \mbox{\Big
)}$.
\end{tabular}

\noindent
In fact only the third and fourth generations contribute here in
the sums.

\newpage
\begin{figure}[t]
{\epsfxsize=200mm \epsfbox[75 0 475 250]{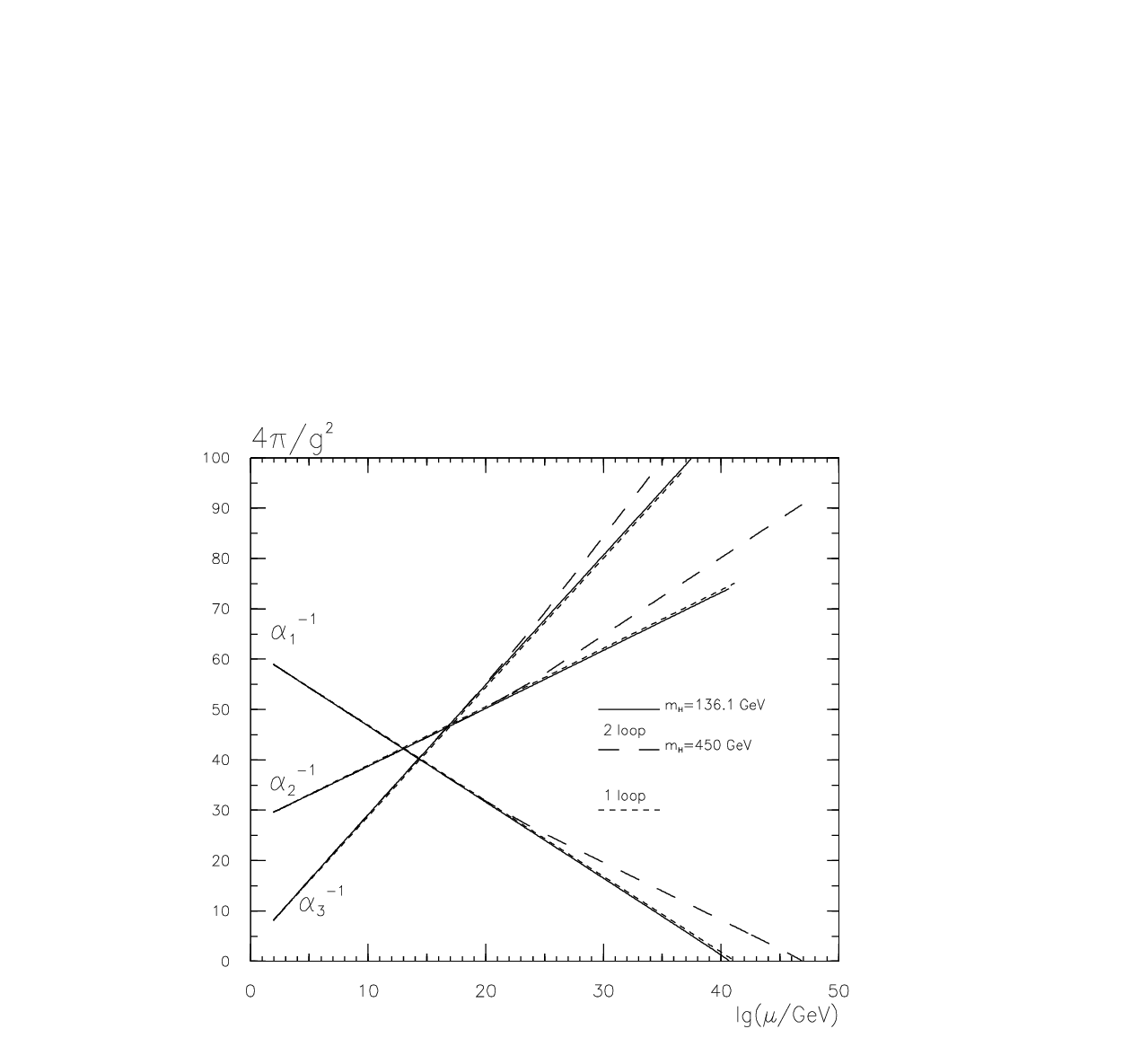}}
\paragraph{}
\paragraph{Fig.~1:~}{
Running of the inverse gauge couplings squared
$\alpha_i^{-1}$,~$i=1,2,3$.
Number of generations is $n_g=3$. Represented Higgs masses are those
corresponding to the typical heavy Higgs and to the lower 
critical Higgs curve  shown in bold in Fig.~3.
}
\end{figure}

\newpage
\begin{figure}[t]
{\epsfxsize=175mm \epsfbox[59 0 459 340]{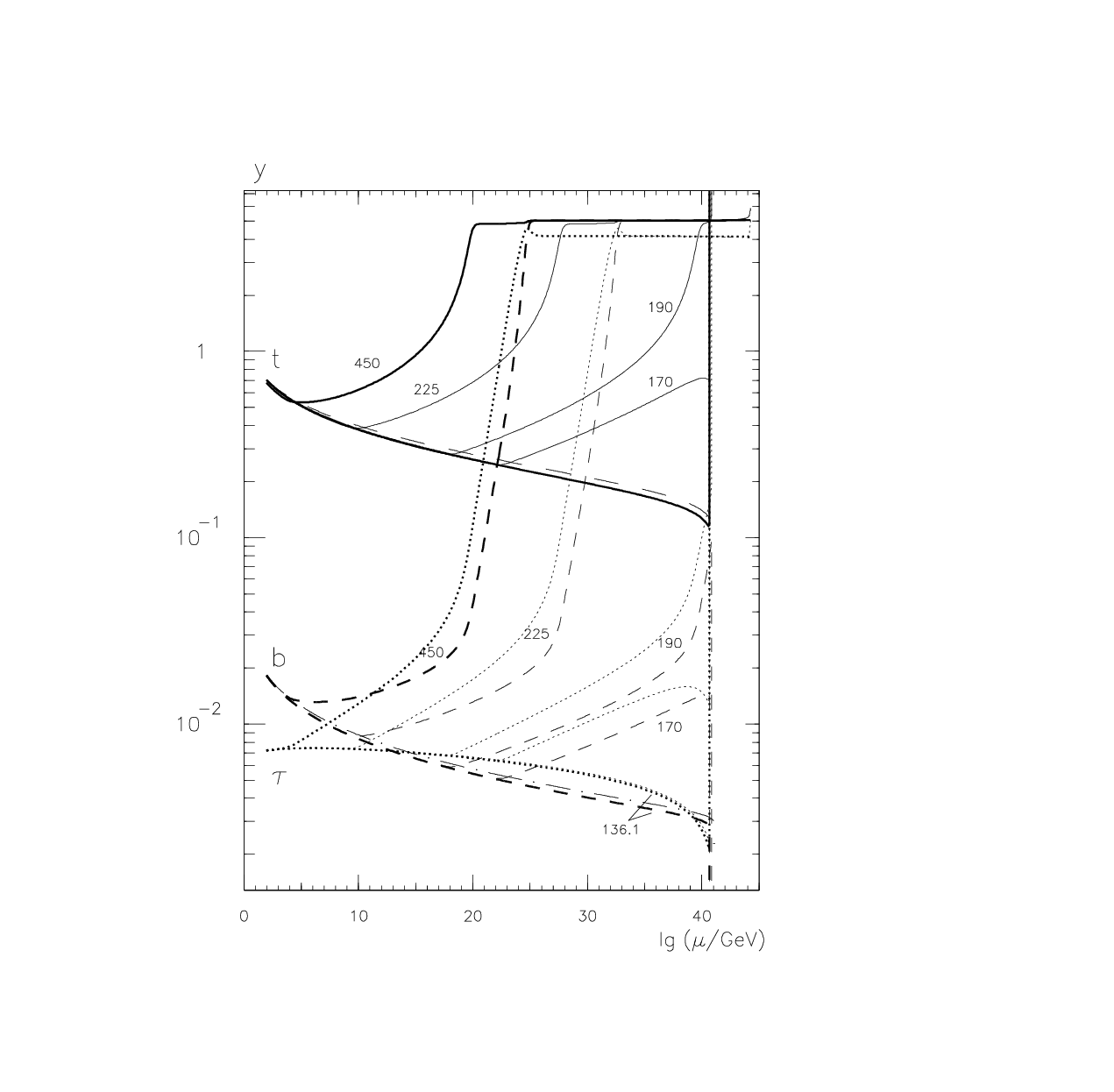}}
\paragraph{Fig.~2:~}{
Running of the third family Yukawa couplings ($n_g=3$).
The falling down curves shown in bold correspond to the lower
critical Higgs mass. The thin lines, close to the latter bold ones,
correspond to the one-loop approximation.  
}
\end{figure}

\newpage
\begin{figure}[t]
{\epsfxsize=200mm \epsfbox[54 0 454 340]{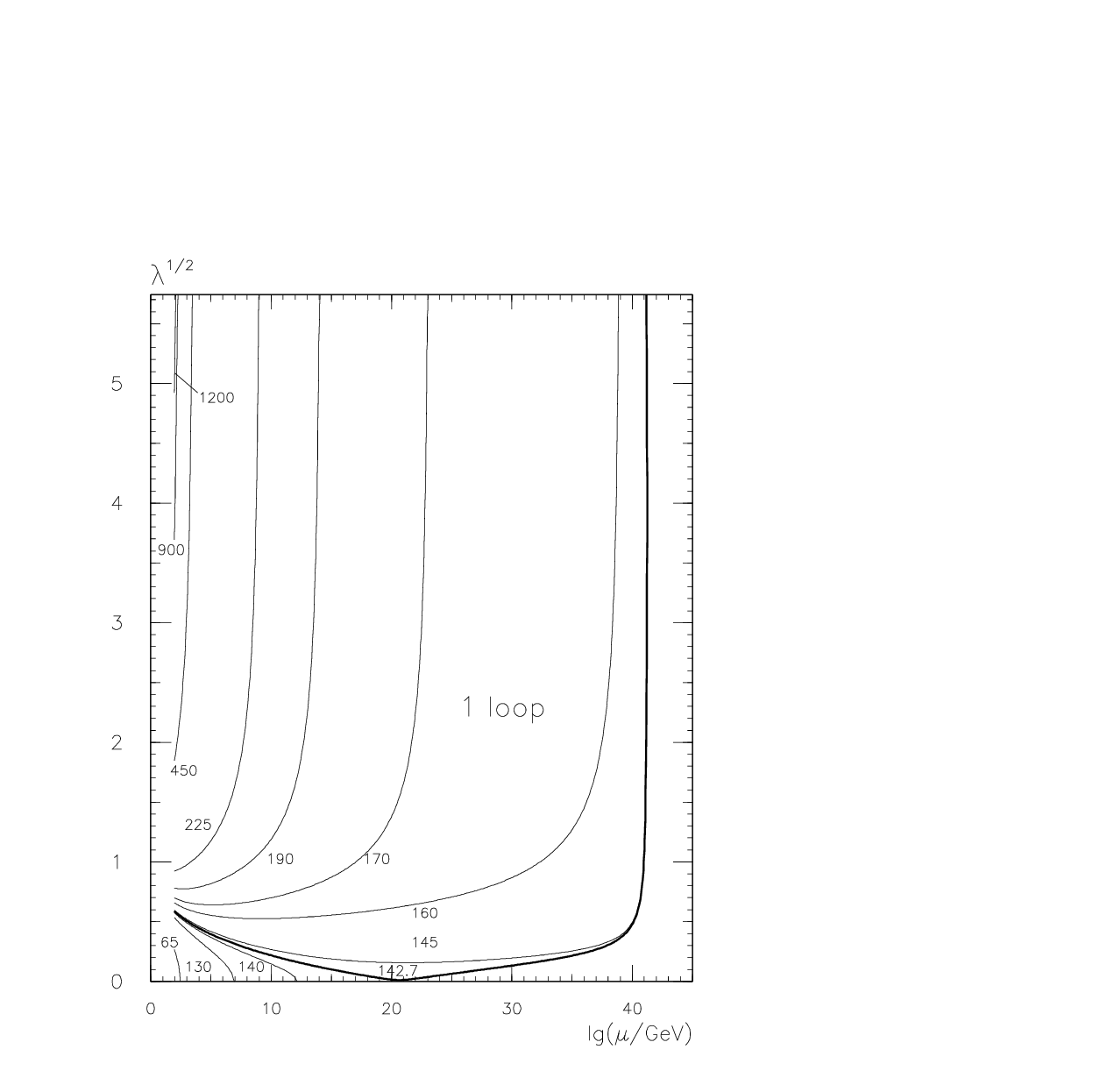}}
\paragraph{Fig.~3a:~}{
Running of the Higgs quartic coupling ($n_g=3$) in one loop.
The critical curves are shown in bold.
}
\end{figure}

\newpage
\begin{figure}[t]
{\epsfxsize=200mm \epsfbox[62 0 462 340]{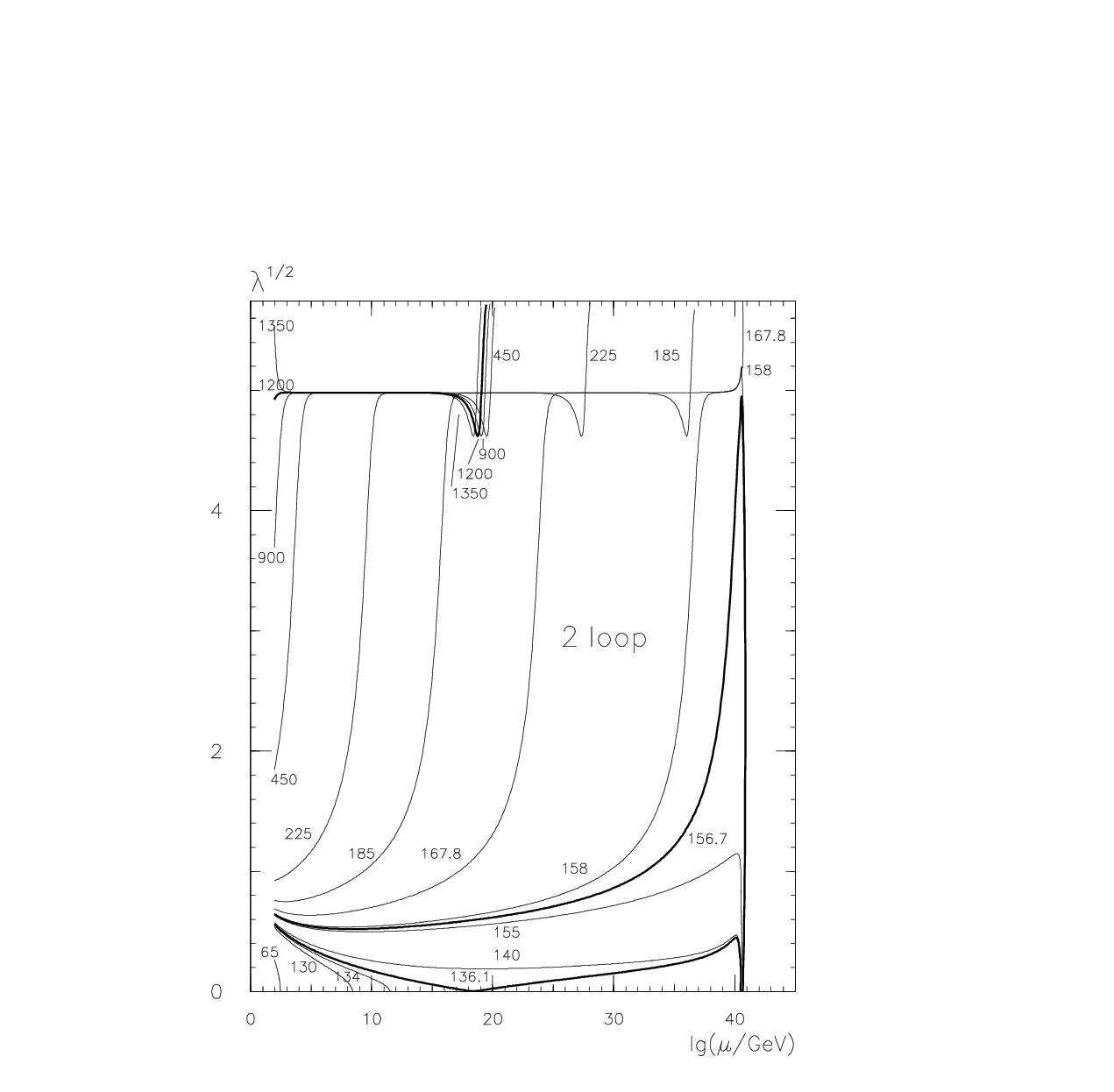}}
\paragraph{Fig.~3b:~}{
Running of the Higgs quartic coupling ($n_g=3$) in two loops.
The rest is as in Fig.~3a.
}
\end{figure}

\newpage
\begin{figure}[t]
{\epsfxsize=150mm \epsfbox[65 0 465 340]{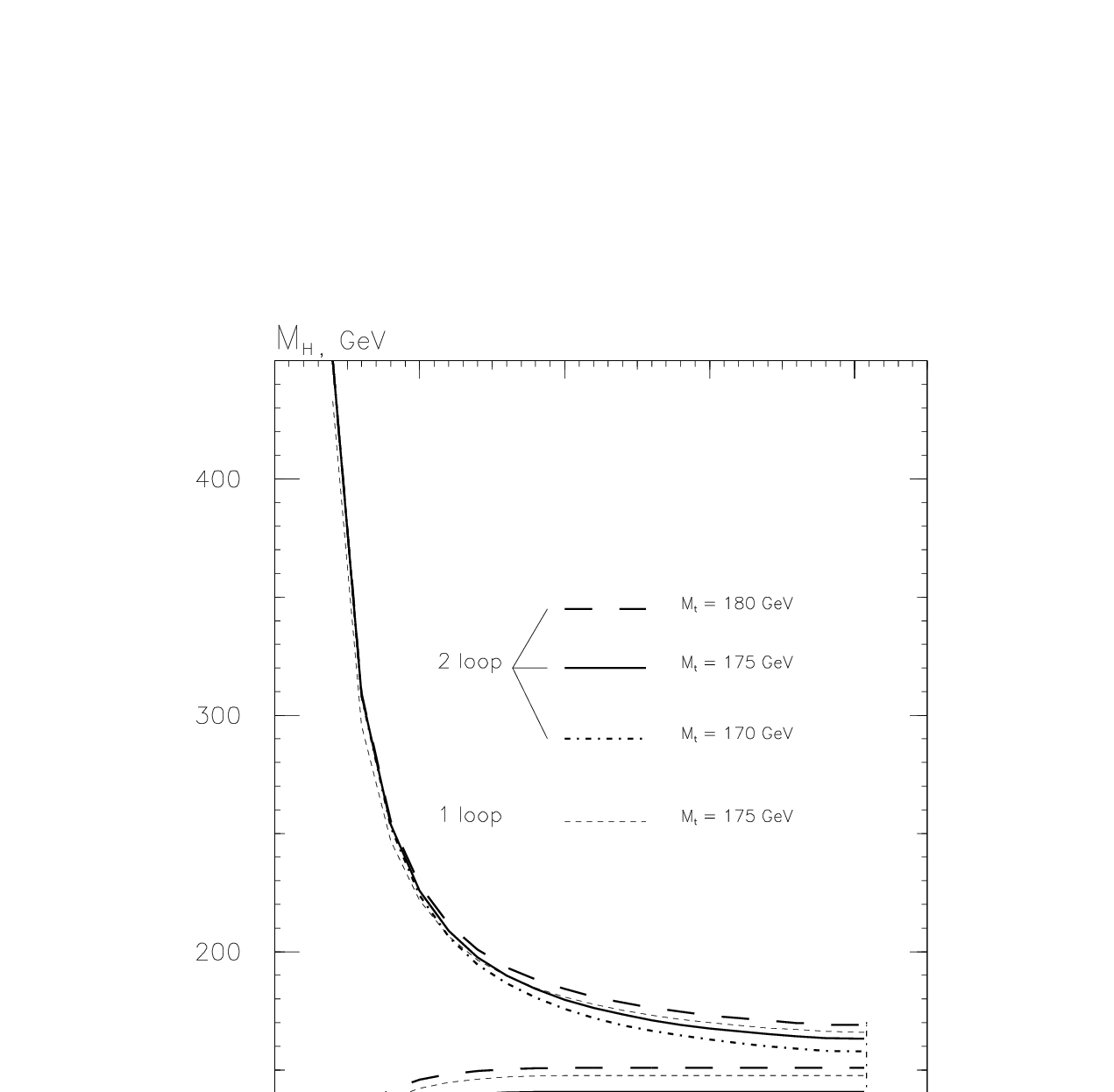}}
\paragraph{Fig.~4:~}{
The SM one- and two-loop  self-consistency plot ($n_g=3$): the
allowed Higgs mass vs.\ the cutoff scale~$\Lambda$.
}
\end{figure}

\newpage
\begin{figure}[t]
{\epsfxsize=200mm \epsfbox[54 30 454 380]{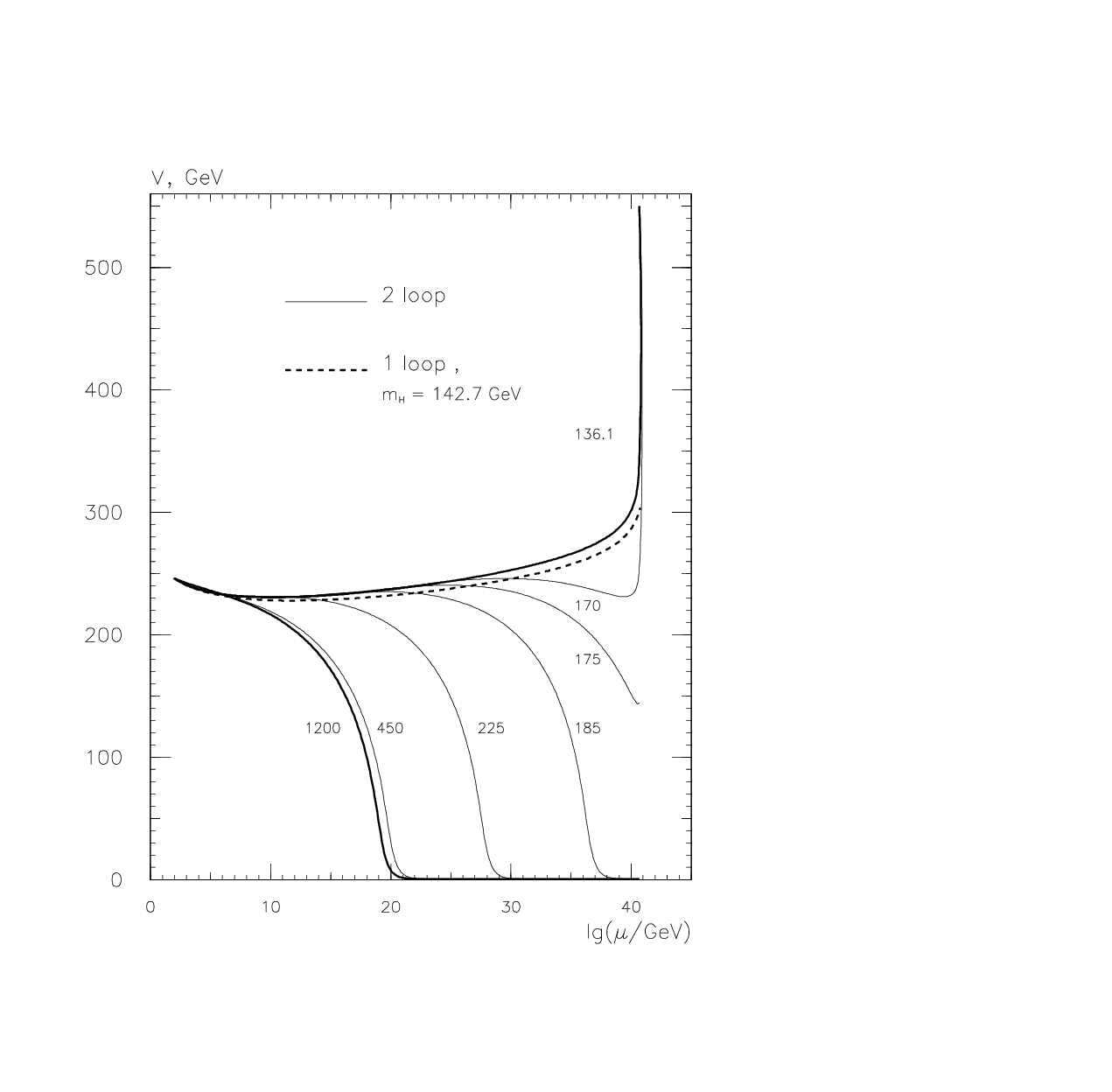}}
\paragraph{Fig.~5:~}{
Running of the  Higgs VEV ($n_g=3$)  in the 't~Hooft--Landau gauge.
}
\end{figure}

\newpage
\begin{figure}[t]
{\epsfxsize=200mm \epsfbox[75 0 475 250]{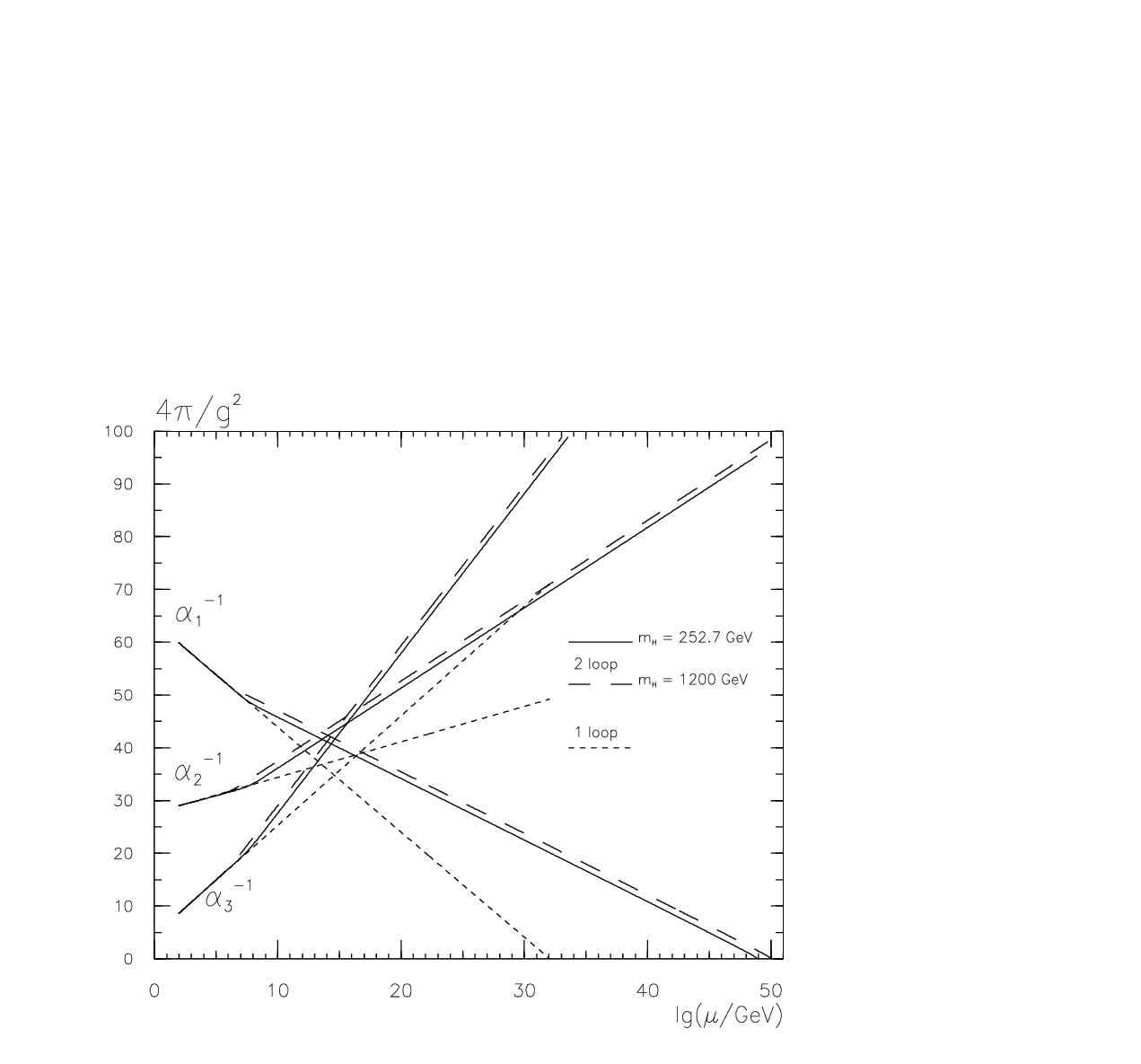}}
\paragraph{Fig.~6a:~}{
Running of the gauge couplings ($n_g = 4$).
The fourth family mass scale is $m_4 = 200$ GeV and $m_L/m_Q = 1/2$.
}
\end{figure}

\newpage

\begin{figure}[t]
{\epsfxsize=200mm \epsfbox[75 0 475 250]{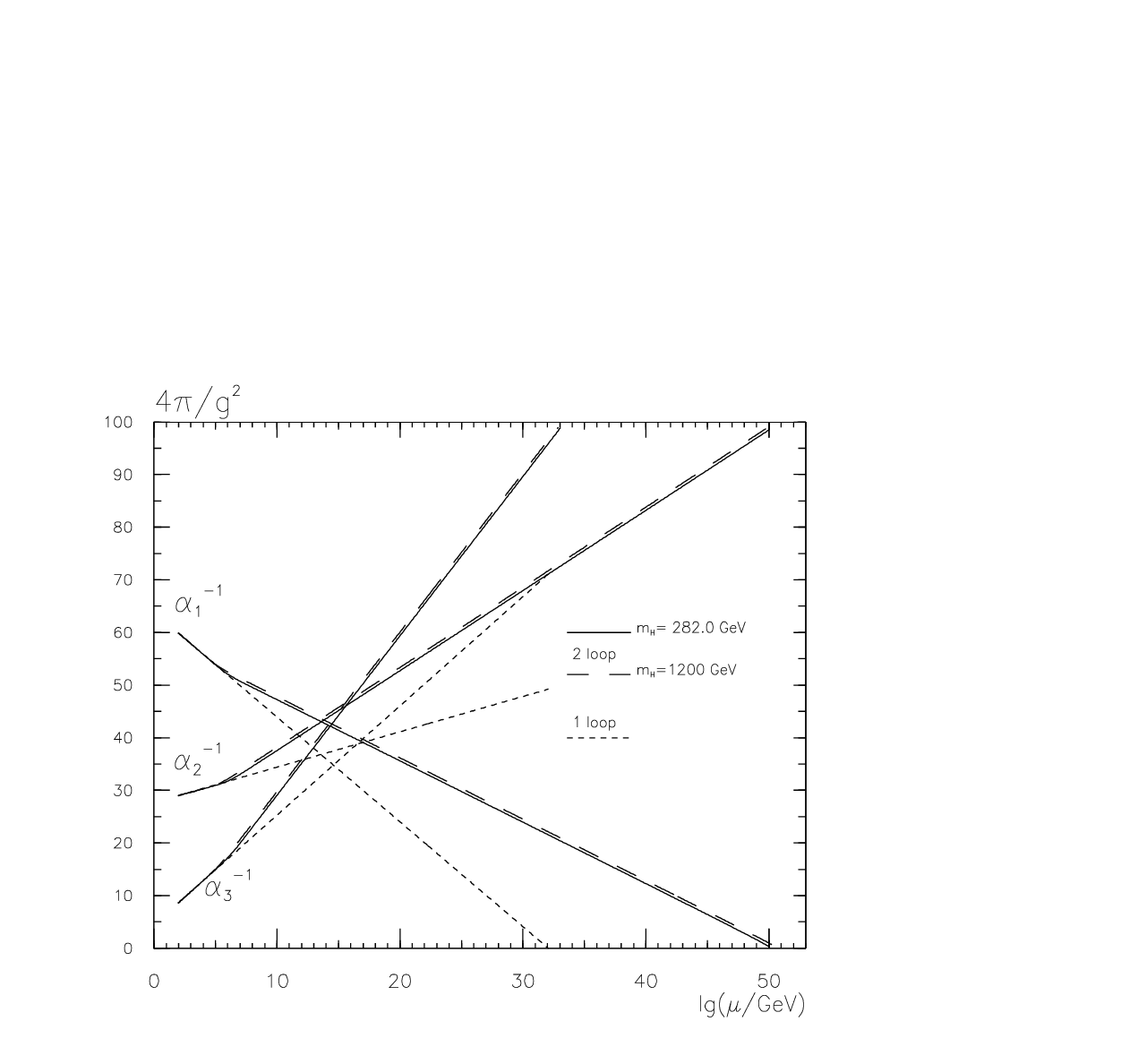}}
\paragraph{Fig.~6b:~}{
The same as in Fig.~6a at  $m_L/m_Q = 1$.
}
\end{figure}

\newpage
\begin{figure}[t]
{\epsfxsize=175mm \epsfbox[40 -20 440 340]{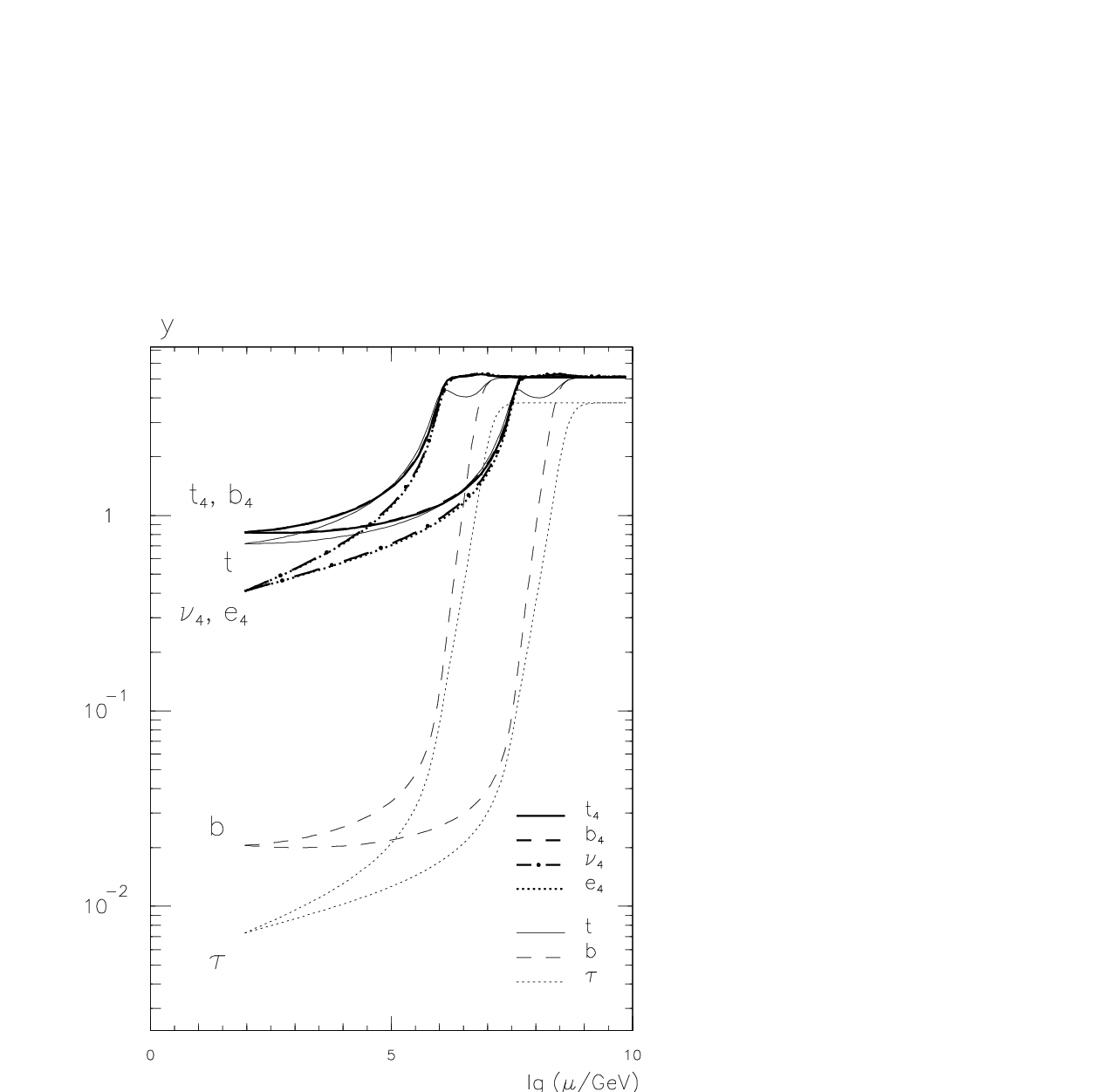}}
\paragraph{Fig.~7a:~}{
Two-loop running of the Yukawa couplings ($n_g = 4$) for the third
and fourth families at $m_4 = 200$ GeV and $m_L/m_Q = 1/2$.
The upper and lower curves correspond to the Higgs masses,
respectively, for the upper and lower Higgs critical curves shown in
bold in Fig.~8.
}
\end{figure}

\newpage
\begin{figure}[t]
{\epsfxsize=175mm \epsfbox[40 -20 440 340]{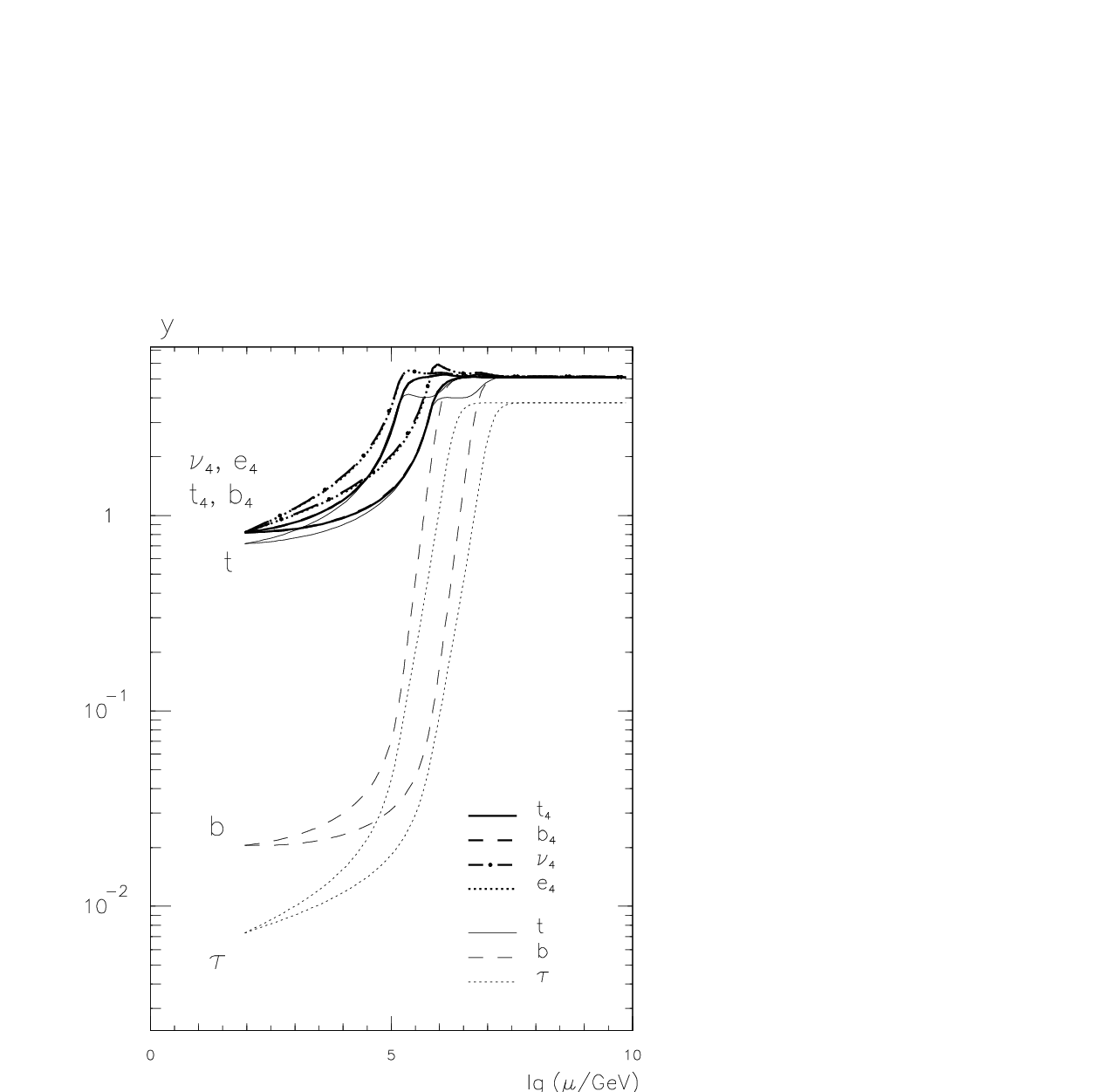}}
\paragraph{Fig.~7b:~}{
The same as in Fig.~7a at  $m_L/m_Q = 1$.
}
\end{figure}

\newpage
\begin{figure}[t]
{\epsfxsize=200mm \epsfbox[40 0 440 300]{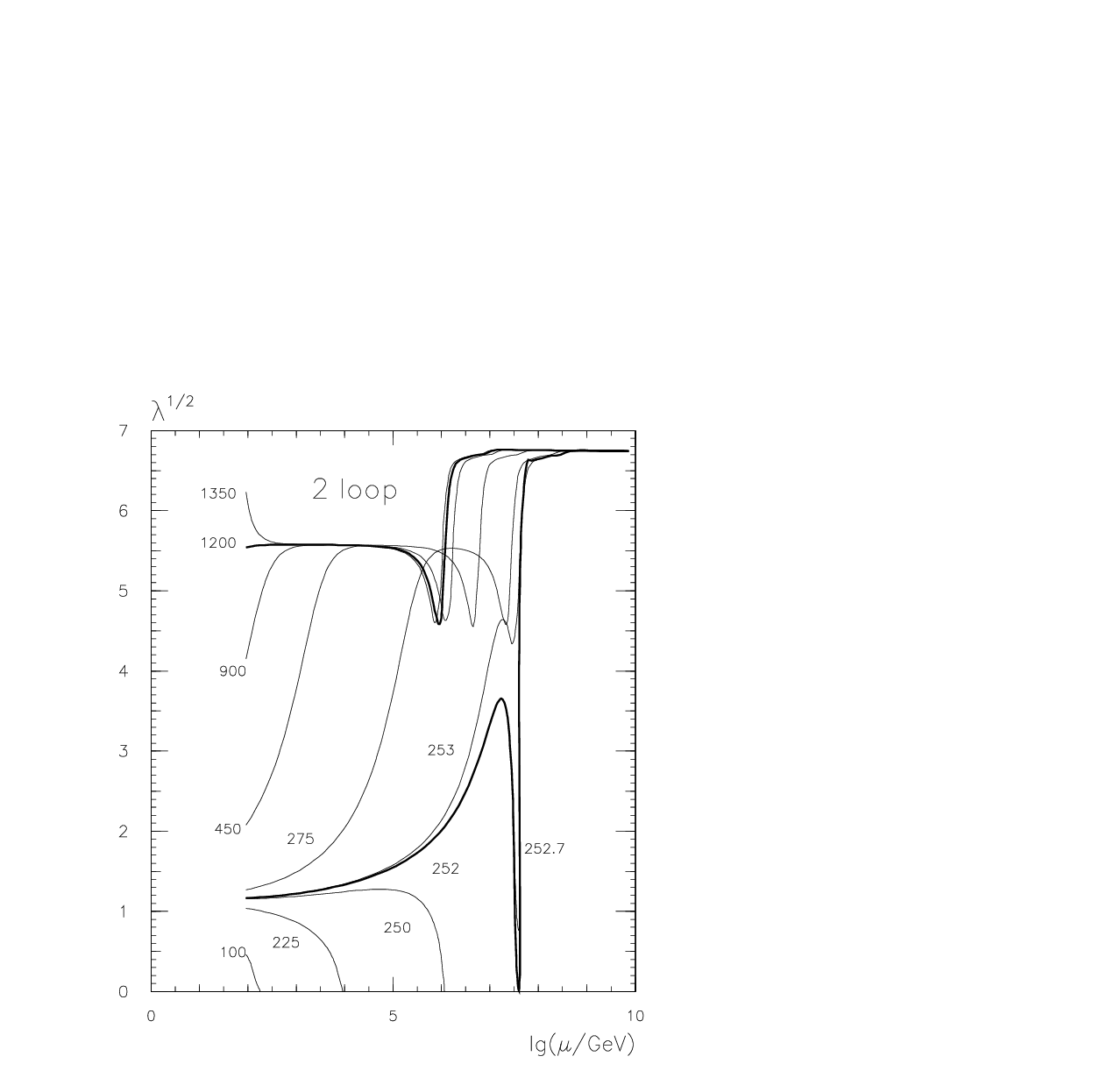}}
\paragraph{Fig.~8a:~}{
Two-loop running of the Higgs quartic coupling ($n_g=4$) at   
$m_4 = 200$ GeV and  $m_L/m_Q = 1/2$.
}
\end{figure}

\newpage
\begin{figure}[t]
{\epsfxsize=200mm \epsfbox[40 0 440 300]{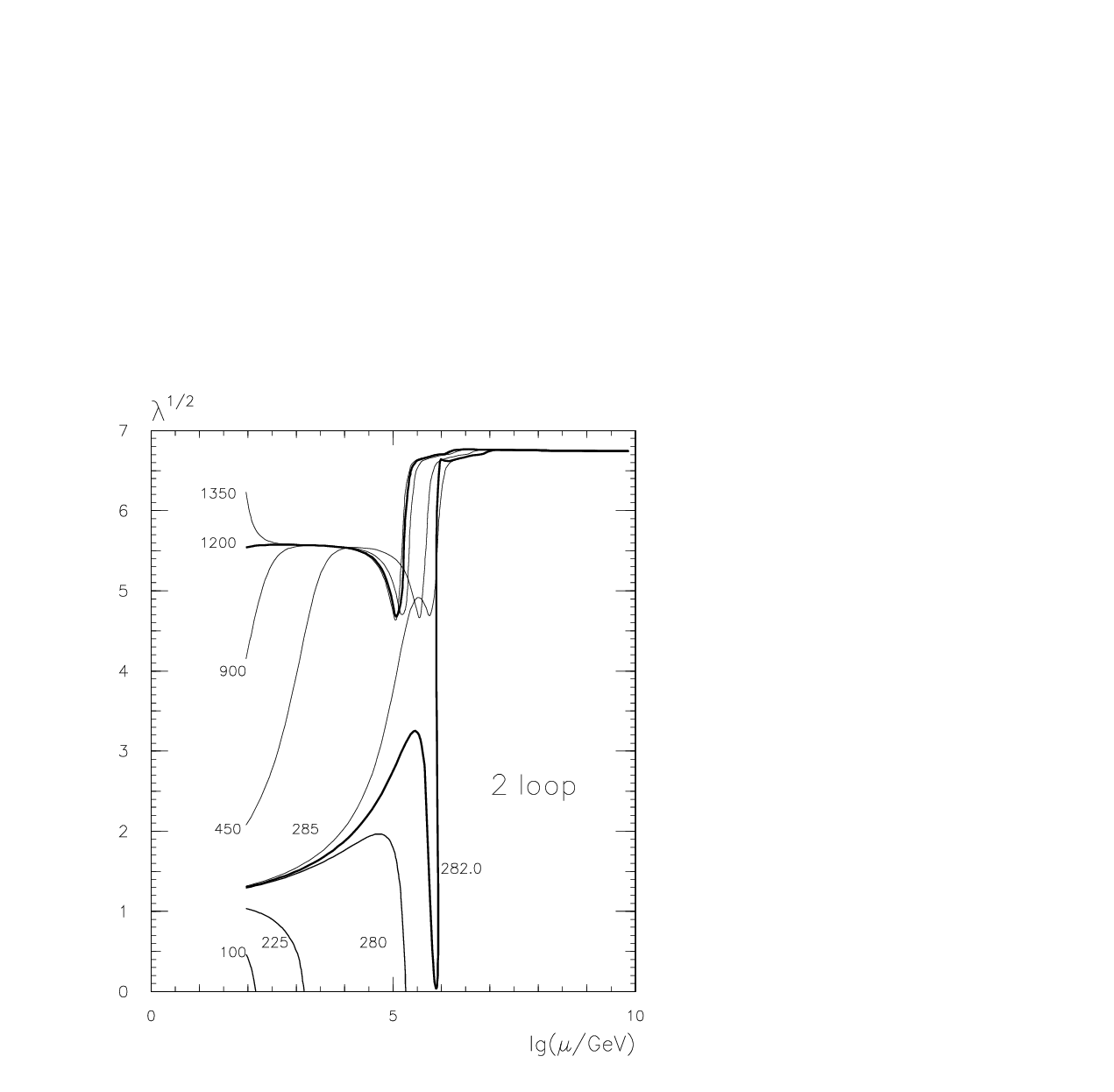}}
\paragraph{Fig.~8b:~}{
The same as in Fig.~8a at  $m_L/m_Q = 1$. 
}
\end{figure}

\newpage
\begin{figure}[t]
{\epsfxsize=150mm \epsfbox[40 0 440 340]{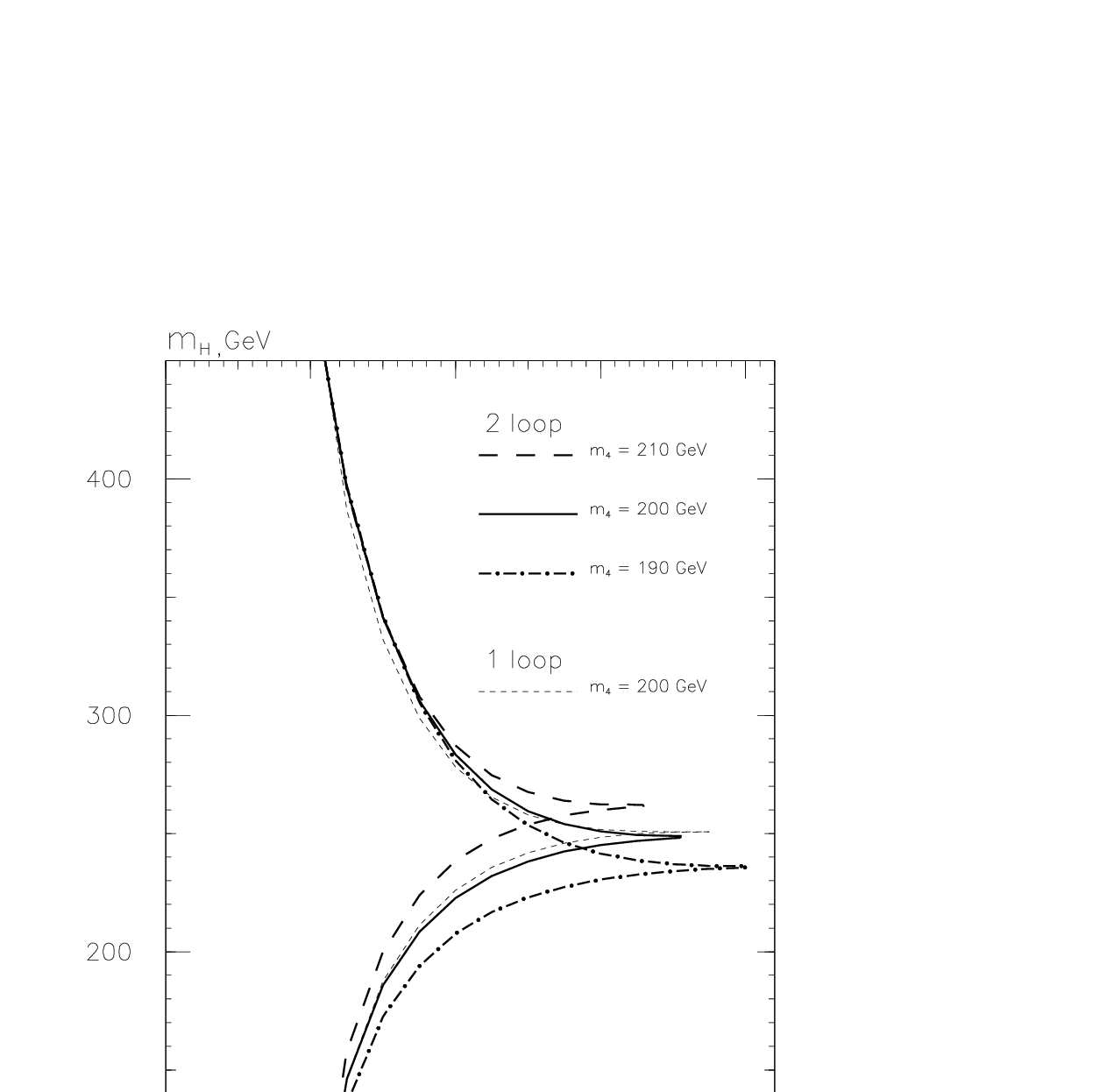}}
\paragraph{Fig.~9a:~}{
One- an two-loop self-consistency plot ($n_g=4$): the allowed Higgs
mass vs.\ the cutoff scale $\Lambda$  at $m_4 = 200$ GeV and $m_L/m_Q
=
1/2$.
}
\end{figure}

\newpage
\begin{figure}[t]
{\epsfxsize=150mm \epsfbox[40 0 440 340]{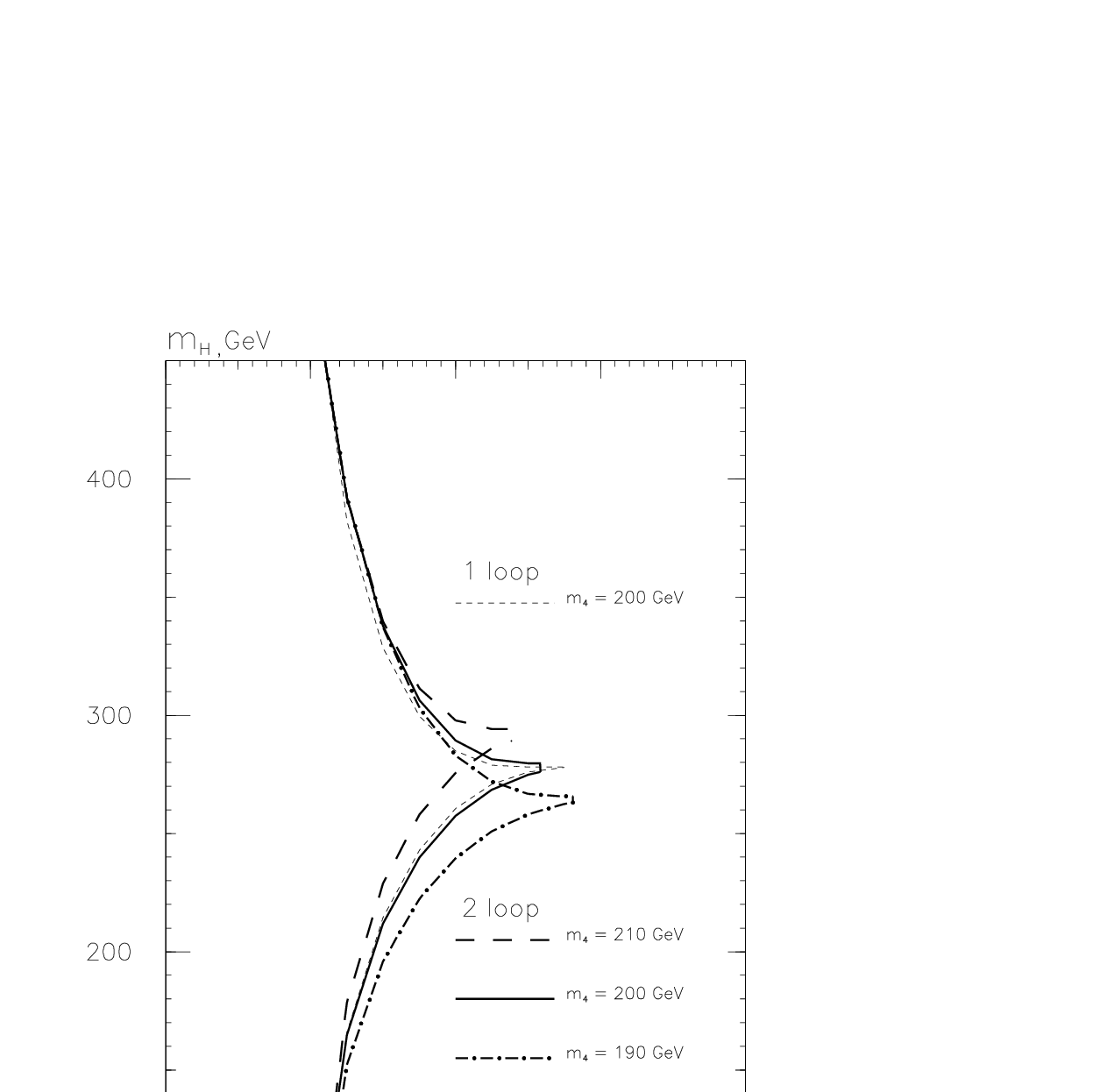}}
\paragraph{Fig.~9b:~}{
The same as in Fig.~9a at  $m_L/m_Q = 1$.
}
\end{figure}

\newpage
\begin{figure}[t]
{\epsfxsize=200mm \epsfbox[40 0 440 300]{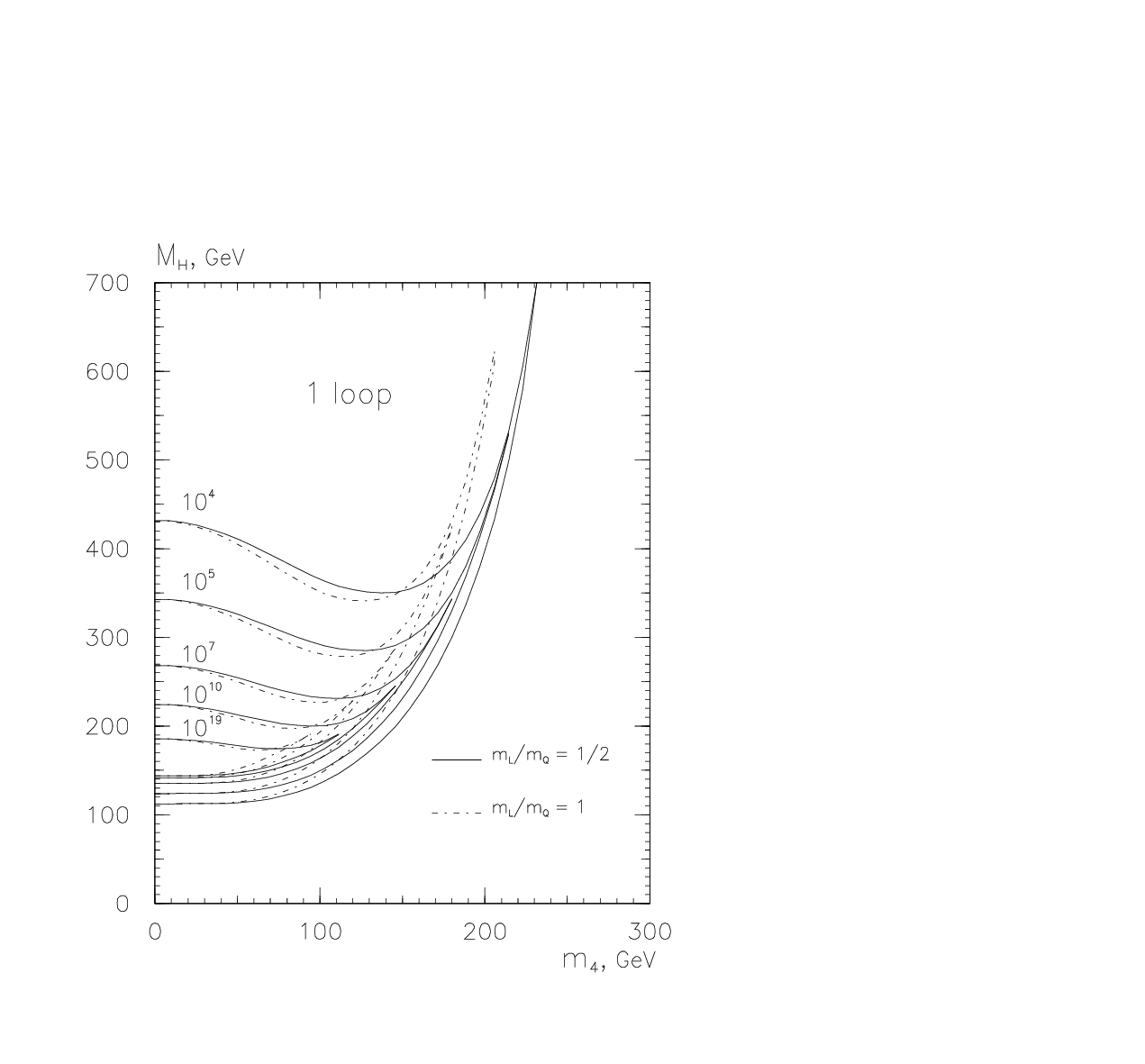}}
\paragraph{Fig.~10:~}{
One-loop self-consistency plot ($n_g=4$): the allowed Higgs mass
vs.\ the fourth family scale $m_4$.  The cutoff
scale $\Lambda$ in GeV is fixed and $M_t=175$ GeV. 
}
\end{figure}

\newpage
\begin{figure}[t]
{\epsfxsize=200mm \epsfbox[40 0 440 300]{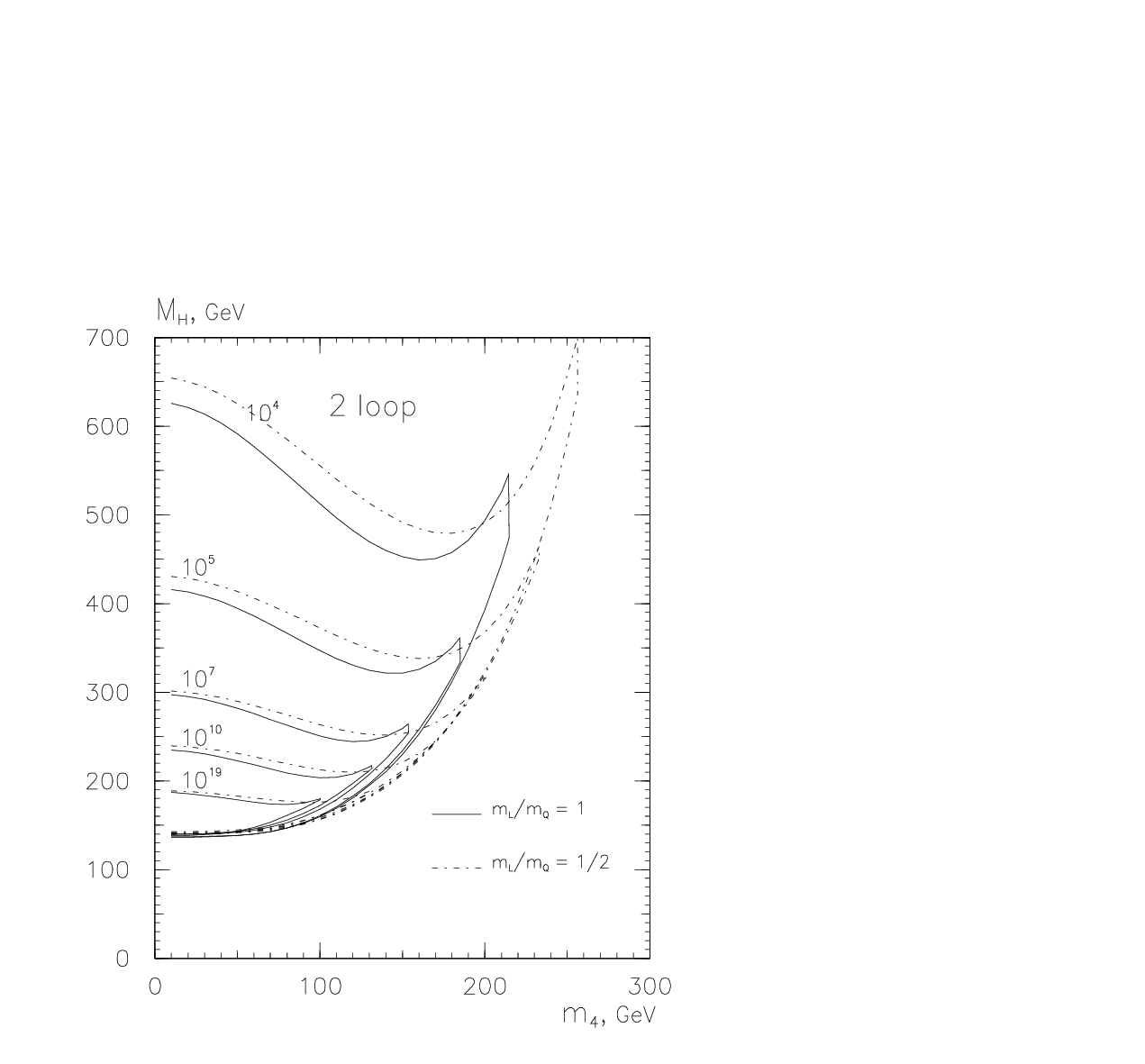}}
\paragraph{Fig.~11a:~}{
Two-loop self-consistency plot  under the restriction $y\leq 1.5$
on the Yukawa couplings ($n_g=4$). The rest is as in Fig.~10. 
}
\end{figure}

\newpage
\begin{figure}[t]
{\epsfxsize=200mm \epsfbox[40 0 440 300]{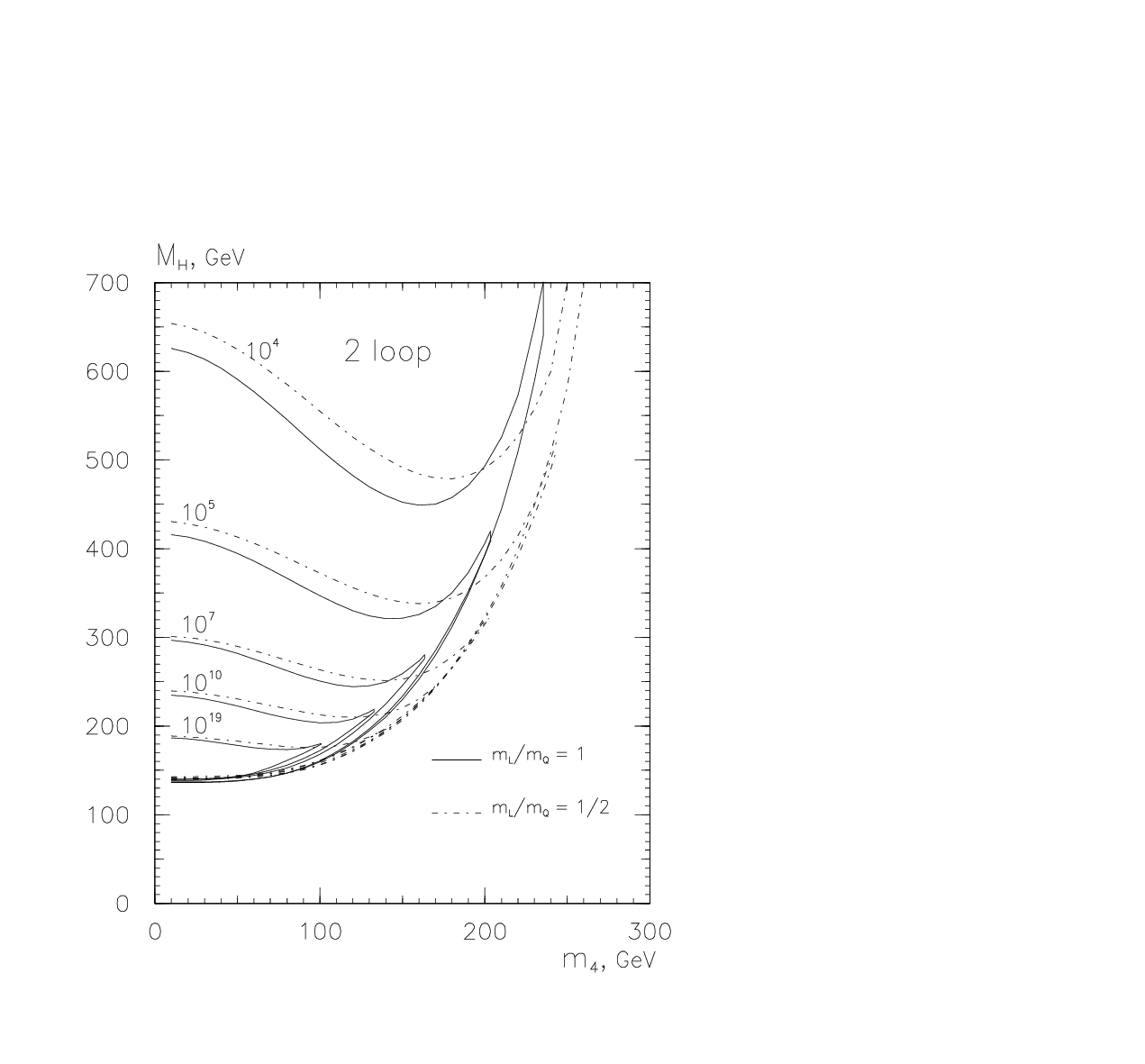}}
\paragraph{Fig.~11b:~}{
The same as in Fig.~11a at  $y\leq 2$.
}
\end{figure}

\end{document}